\documentclass[preprint,12pt,authoryear]{elsarticle}

\usepackage{amssymb}
\usepackage{amsmath}
\usepackage{wasysym}
\usepackage{longtable}
\usepackage{tabu}
\usepackage{setspace}
\usepackage{lscape}
\usepackage{fancyhdr}
\usepackage{fullpage}
\usepackage{caption}
\usepackage{gensymb}
\usepackage{outlines}
\usepackage{lineno}
\usepackage{xcolor}
\usepackage{ulem}

\journal{Geochemistry}

\begin{document}

\begin{frontmatter}



\title{Comet 81P/Wild 2: A Record of the Solar System's Wild Youth}


\author{Ryan C. Ogliore}

\affiliation{organization={McDonnell Center for the Space Sciences, Washington University in St.\ Louis},
            addressline={1 Brookings Dr.}, 
            city={St. Louis},
            postcode={63130}, 
            state={MO},
            country={USA}}

\begin{abstract}

\end{abstract}




\end{frontmatter}

\tableofcontents

\section*{Abstract}
NASA's Stardust mission returned rocky material from the coma of comet 81P/Wild~2 (pronounced ``Vilt 2'') to Earth for laboratory study on January 15, 2006. Comet Wild~2 contains volatile ices and likely accreted beyond the orbit of Neptune. It was expected that the Wild~2 samples would contain abundant primordial molecular cloud material---interstellar and circumstellar grains. Instead, the interstellar component of Wild~2 was found to be very minor, and nearly all of the returned particles formed in broad and diverse regions of the solar nebula. While some characteristics of the Wild~2 material are similar to primitive chondrites, its compositional diversity testifies to a very different origin and evolution history than asteroids. Comet Wild~2 does not exist on a continuum with known asteroids. Collisional debris from asteroids is mostly absent in Wild~2, and it likely accreted dust from the outer and inner Solar System (across the putative gap created by a forming Jupiter) before dispersal of the solar nebula. Comets are a diverse set of bodies, and Wild~2 may represent a type of comet that accreted a high fraction of dust processed in the young Solar System.

\section{Introduction}
\label{intro}
Comets are small Solar System bodies that accreted in cryogenic conditions in the outer Solar System. By definition, comets show ``cometary activity'' as they near the Sun: ejection of gas and dust due to ice sublimation which forms a coma and dust tail. Comets contain a substantial fraction of ice but are still mostly made of rocky material (``frosty rocks'' instead of ``dirty snowballs'') \citep{kuppers2005large}. The surfaces of comets are shaped by ice sublimation which results in the loss of about 25\% of their diameters during their active lifetimes \citep{thomas2009nuclei}. This contrasts with the surfaces of asteroids, which have old surfaces covered in impact craters. The retention of ices in cometary nuclei means they have not been heated like planets and asteroids.  Cometary dust should, therefore, be unmodified since its accretion into the nucleus.


\subsection{Short- and Long-period Comets}

Comets can be classified as long-period or short-period. Long-period comets have random orbital inclinations and orbital periods longer than 200 years; short-period comets have orbital periods shorter than 200 years. Of the short-period comets, Jupiter-family comets (JFCs) have orbital periods less than 20 years and Halley-type comets have periods greater than 20 years. The Tisserand parameter with respect to Jupiter, $T_J$, is defined as:
\begin{equation*}
T_J = \frac{a_J}{a} + 2\cos{(i)}\left[ (1-e^2) \frac{a}{a_J} \right]^{1/2}
\end{equation*}
with $a$ the semi-major axis of the object, $e$ its eccentricity, $i$ its inclination, and $a_J$ Jupiter's semi-major axis. Close encounters with Jupiter change a body's orbital parameters, but $T_J$ stays approximately constant. Jupiter-family comets, which originated beyond Jupiter's orbit, have $T_J$ values between 2 and 3 \citep{levison1997kuiper}. Regular asteroids, which originated inside Jupiter's orbit, have $T_J$ greater than 3. Jupiter-family comets also have low-inclination prograde orbits with aphelia between 5 and 6 AU. The orbits of JFCs are chaotic on long timescales due to interactions with Jupiter at 5.2 AU \citep{lowry2008kuiper}.  

Long-period comets are thought to originate in the Oort cloud, a roughly spherical distribution of icy bodies up to 5$\times$10$^4$ AU from the Sun with a total mass approximately equal to the mass of Earth. Objects in the Oort cloud are too distant to be directly observed but must exist to explain the observed comets with large aphelia. The extent of the Oort cloud, more than 1000 times farther from the Sun than the planets, indicates that these objects did not form in place as the density of material in the protoplanetary disk at these distances was far too low. Oort cloud comets likely were scattered to distant orbits by Jupiter and Saturn \citep{weissman1999diversity}. The outer edge of the Oort cloud is defined by loss due to gravitational interactions with, e.g., passing stars.

Short-period comets were first thought to be Oort-cloud comets that were perturbed into smaller orbits after an interaction with Jupiter. \citet{fernandez1980existence} showed that the Oort cloud could not produce enough comets to match observations, and instead short-period comets likely originated from a disk-like region of icy bodies beyond the orbit of Neptune. Dynamical studies have shown that scattered disk objects (SDOs) are likely the source of most JFCs \citep{duncan1997disk}. SDOs themselves likely originated in the inner Kuiper belt or in the Uranus-Neptune region \citep{lowry2008kuiper}. Kuiper Belt objects (KBOs) have been observed up close by the New Horizons mission \citep{keane2022geophysical}. The Kuiper Belt contains many large ($>$ 50 km) bodies (more than are found in the asteroid belt), the largest of which are the dwarf planets Eris and Pluto.


\subsection{Remote-Sensing Measurements of Comets, Protoplanetary Disks, and the Interstellar Medium}

The volatile content of comets has been measured for several long-period and short-period comets. At least some species appear to be highly variable. The cometary activity of the JFC Hartley 2 is mostly driven by CO$_2$ release, as measured by NASA's EPOXI mission \citep{a2011epoxi}. The release of super-volatile CO$_2$ carries out water ice, which sublimes to make water vapor in the comet's coma. Another super-volatile, CO, is much less abundant than CO$_2$ in Hartley 2 \citep[CO$_2$/CO $>$ 60][]{weaver2011carbon}. This contrasts with the measured ratio in the JFC Tempel 1 \citep[CO$_2$/CO $\approx$1, ][]{feaga2007comparison}, short-period comet Halley \citep[$<$1, ][]{bockelee2004composition}, and long-period comet Hale-Bopp  \citep[$\sim$1, ][]{bockelee2003lessons}.  Some JFCs are highly depleted in carbon-chain molecules \citep{a1995ensemble}. In nine non-carbon-depleted JFCs, \citet{a1995ensemble} measured the production rate ratios of C$_2$/OH, C$_3$/OH, CN/OH, NH/OH, to vary by factors of 3, 8, 3, and 4, respectively.  KBOs are observed to vary greatly in their spectral reflectance: from neutrally reflecting to some of the Solar System's reddest bodies \citep[e.g.,][]{luu1996color}. This is possibly due to radiation processing of volatiles such as H$_2$O, H$_2$S, CO$_2$, CH$_3$OH, and NH$_3$. The relative abundance of volatile species varies as a function of distance in the Kuiper Belt \citep{parhi2023sublimation}, so the observed color differences in KBOs could reflect differences in the relative abundances of volatiles. 

Hydrogen isotopes measured spectroscopically in cometary water show enormous variability. Both Oort-cloud and Jupiter-family comets show D/H ratios that range between the ratio found in Earth's ocean water \citep{hartogh2011ocean,lis2013herschel,biver2016isotopic} and 3 times this value \citep{bockelee2015cometary,altwegg201567p}. The large relative mass difference between deuterium and hydrogen cause large isotopic variations in natural systems \citep{criss1999principles}. \citet{lis2019terrestrial} proposed that the observed D/H variability in water outgassed from comets may be due to fractionation effects during sublimation. Alternatively, the D/H variability may reflect different reservoirs of water with contributions from the very low protosolar value \citep{geiss1998abundances} and very high values caused by low-temperature chemistry in the interstellar medium and protosolar molecular cloud \citep{aikawa1999deuterium}.

\begin{figure}[!ht]
\begin{center}
\includegraphics[height=0.2\textheight]{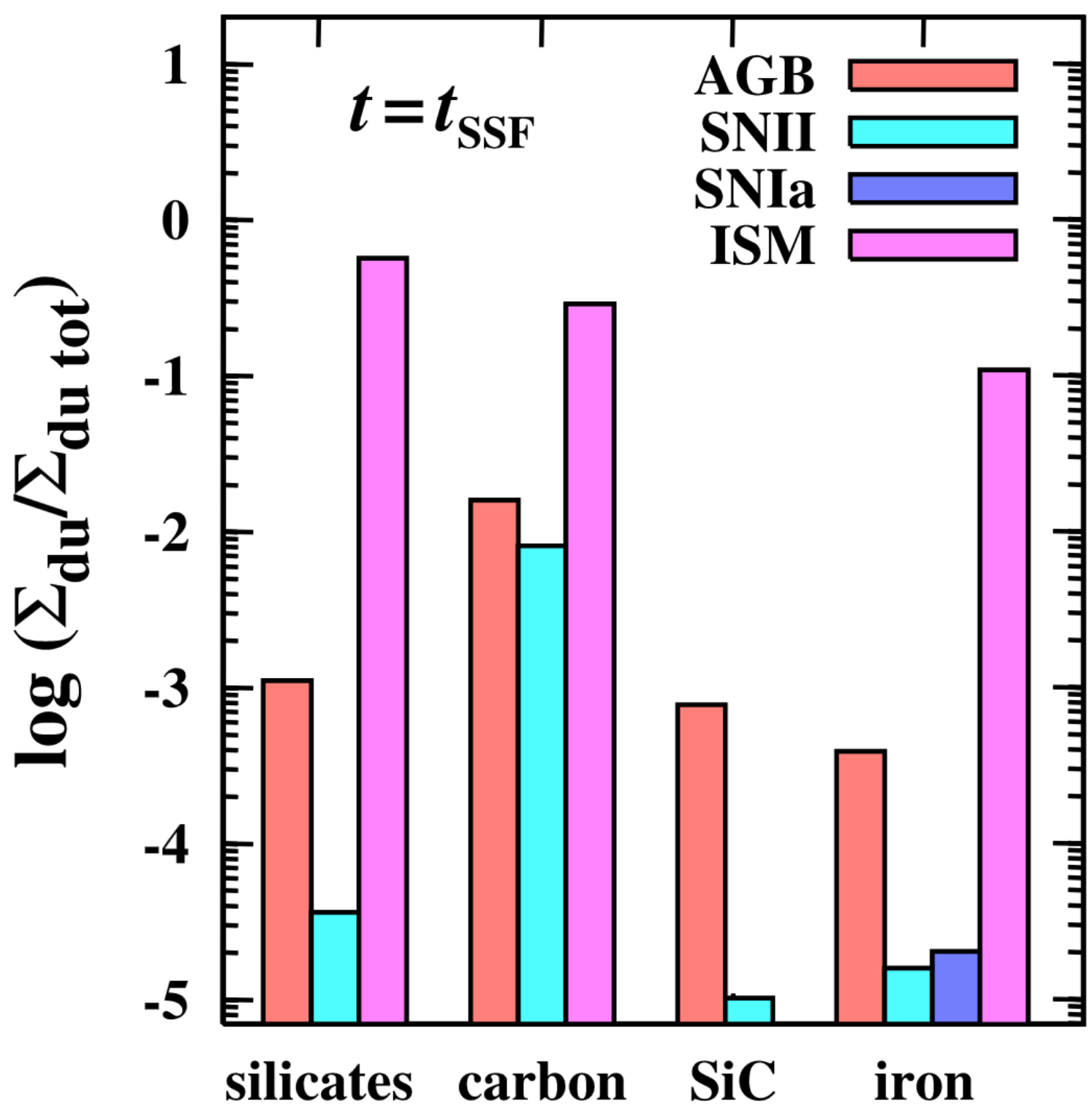}
\includegraphics[height=0.2\textheight]{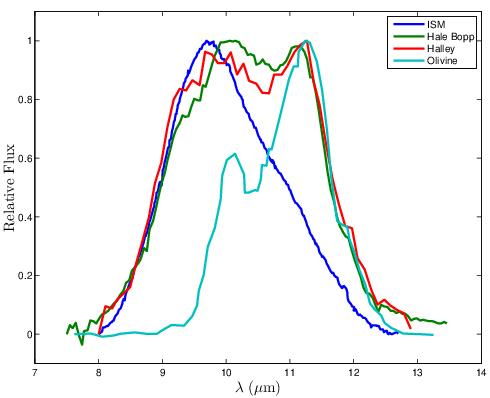}
\includegraphics[height=0.2\textheight]{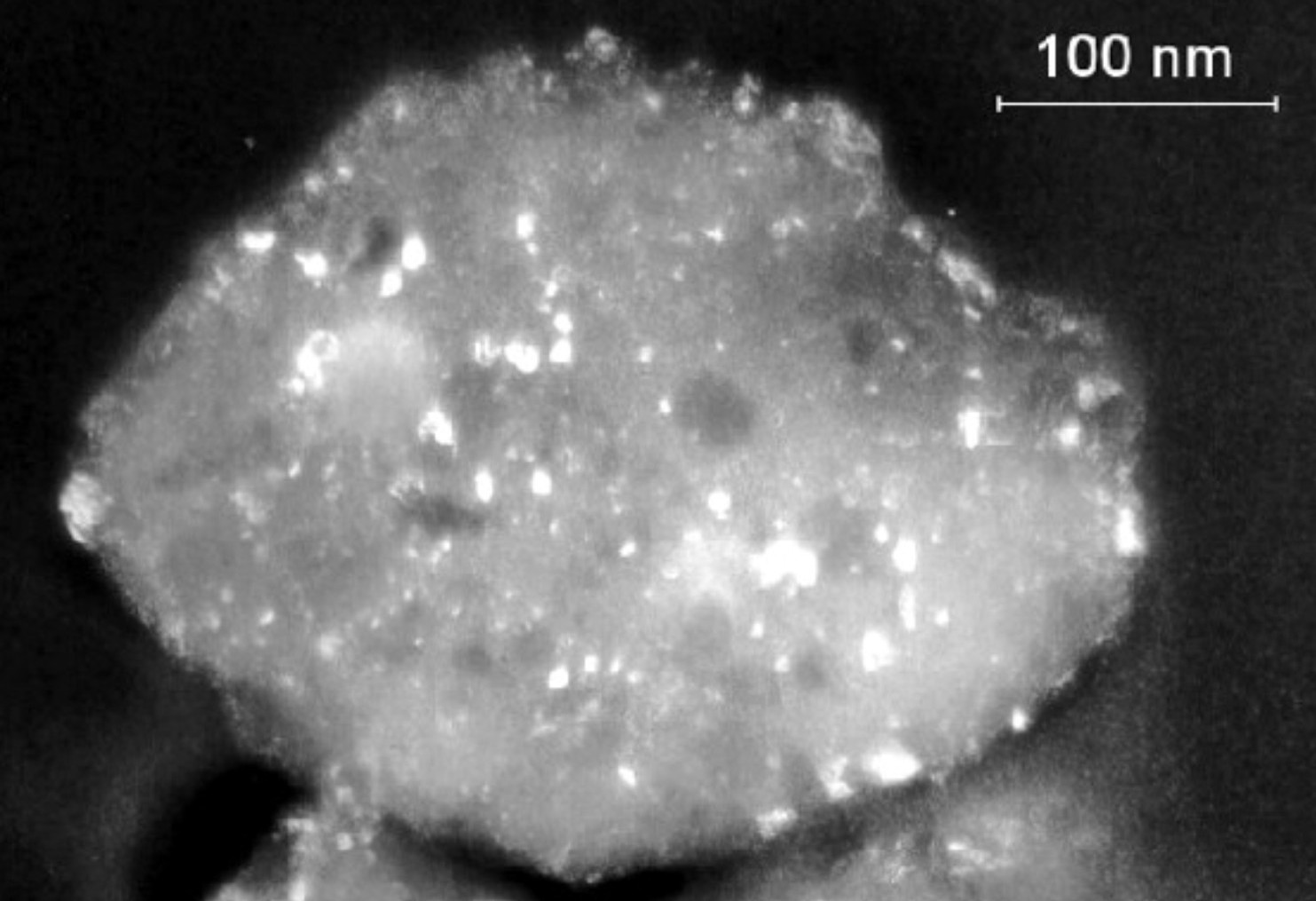}
\end{center}
\vspace*{-0.25in}
\caption{Interstellar dust in models, observations, and samples. Left) Composition of interstellar dust in the solar neighborhood at the time of Solar System formation, from \citet{zhukovska2008evolution}. Center) Infrared spectrum around 10 $\mu$m of the interstellar medium \citep{kem04}, comet Hale Bopp \citep{hanner2004composition}, comet Halley \citep{hanner19948}, and olivine powder \citep{stephens1979emission}. All spectra are scaled for comparison. Right) Dark-field TEM image of a GEMS grain (candidate interstellar amorphous silicate) from \citep{bradley1999infrared} showing metal and sulfides (bright inclusions) embedded in a Mg-rich amorphous matrix. \label{fig:silicatefeature}}
\end{figure}

The rocky (nonvolatile) component has also been measured spectroscopically in several comets. Vibrational mode transitions in silicate minerals are found in the 10 $\mu$m  region. Infrared spectroscopy of comets often reveals a broad emission feature near 10 $\mu$m, with varying strength, widths, and subpeaks \citep[e.g.,][]{hanner19948}. Many of these features are consistent with amorphous silicates. A small subpeak at 11.1 $\mu$m is associated with the mineral olivine \citep{do2020crystalline}, which is used to distinguish the (crystalline) mineral from amorphous silicates with elemental composition consistent with olivine. The strength, shape, and position of this feature can depend sensitively on the size, shape, and composition of the silicates \citep{hanner2004composition,chornaya2020revisiting}. The 11.1 $\mu$m feature is present in comets Halley \citep{campins1989identification}, Hale-Bopp \citep{hanner1997thermal}, Tempel 1 \citep{harker2005dust}, and others (Figure \ref{fig:silicatefeature}). The crystalline feature is also present in the spectra of pre-main-sequence Herbig Ae/Be stars \citep{waelkens1996}, T-Tauri stars \citep[e.g., ][]{honda2003detection}, and the young main-sequence star Beta Pictoris \citep{knacke1993}. These young stars are surrounded by debris disks of gas and dust, which include crystalline silicates.

The Mg/Fe ratio in cometary crystalline silicates can be estimated from the 11.1 $\mu$m olivine and 9.2 $\mu$m pyroxene feature. This feature is strongest when the emitted dust is 1~$\mu$m or smaller in size. The long-period comet Hale-Bopp had a very strong infrared silicate emission feature, and the relative contributions of Fe-rich and Mg-rich pyroxenes and olivines can be estimated. Hale-Bopp has a very high Mg/Fe ratio in its silicates, dominated by forsterite with contribution from enstatite  \citep{wooden1999silicate}. 

Silicates in the interstellar medium lack this 11.1 $\mu$m feature associated with crystalline olivine (Figure \ref{fig:silicatefeature}).  Infrared observations of young stellar objects, H-II regions, and the interstellar medium show that the crystalline fraction of interstellar silicates is 1--2\% by mass \citep{do2020crystalline}. Molecular clouds also do not show crystalline silicates in their IR spectra \citep{bowey199810}. The fact that crystalline silicates are absent in the interstellar medium and molecular clouds but present in protoplanetary disks and comets implies that some process must create them in the collapse of the molecular cloud, or in the protoplanetary disk itself. Interstellar amorphous silicates could be warmed and annealed to create crystalline material, or olivine could be formed by evaporation and recondensation of amorphous silicates. 

The temperature in the outer solar nebula, where the comets formed, was likely too low to create crystalline silicates by annealing \citep{gail1998chemical}. Additionally, collisions in the Kuiper Belt are relatively low speed and rare, which makes it difficult to convert amorphous ice to crystalline ice \citep{marboeuf2009can}, and also difficult to crystallize amorphous silicates. Very hot temperatures required to vaporize amorphous interstellar dust existed close to the young Sun \citep{neufeld1994}. It is possible that regions of the disk up to $\sim$3~AU from the protostar can be heated to dust-vaporization temperatures during periodic FU-orionis-like outbursts \citep{boss2020evolution}. Some transport process is required to move inner nebula material to the comet-formation region $>$30~AU from the Sun to explain the presence of crystalline silicates in comets. Turbulent diffusion in the solar nebula can accomplish this \citep{bockelee2002turbulent,ciesla2007outward}, but the efficiency of large-scale transport is uncertain as it is model-dependent and varies strongly with viscosity. 

Comets contain crystalline silicates which likely formed in the solar nebula, but they also contain ices which obviously were kept very cold during the lifetime of the comet, as well as amorphous silicates. The rocky material of comets could be made of unprocessed material inherited from the Solar System's parent molecular cloud, with a minor component of inner nebula material transported outward to the comet-forming region. The nonvolatile, rocky material in a comet's coma has a significant component of very small grains, $<$1~$\mu$m, due to the presence of the silicate feature near 10 $\mu$m. Larger particles may be aggregates of smaller grains. Analyses of such small and likely complex particles requires large, complex instruments and cometary samples in Earth laboratories.

\subsection{Dust from comet 67P/Churyumov-Gerasimenko and comet 1P/Halley}

During the last appariton of Halley's Comet in 1986, three spacecraft flew by the comet nucleus and measured impacting dust grains with mass spectrometers. The Vega 1 and Vega 2 spacecraft flew by Halley after studying Venus. Measurements by the PUMA-1 and PUMA-2 instruments showed that Halley dust was a mixture of silicates (with variable Fe/Mg) and organics. The organic/silicate ratio appeared to change between the PUMA-1 and PUMA-2 measurements, indicating that the composition of Halley's coma was time variable \citep{fomenkova1992compositional}. Particles dominated by carbon or organics (with minor silicate component, \citet{lawler1992chon}) make up 22\% of all dust \citep{fomenkova1994carbonaceous} but are more abundant closer to the nucleus. The rest of the measured Halley dust was silicate particles with minor carbon and particles with close to equal mixtures of silicates and carbonaceous material.

Comet 67P/Churyumov-Gerasimenko (C-G) was studied in detail by the Rosetta mission. The MIDAS instrument collected C-G dust particles and imaged them with an atomic-force microscope using a novel reverse-imaging technique \citep{mannel2019dust} with resolution better than 100~nm. The majority of particles measured by MIDAS were large compact agglomerates, though one $>$10~$\mu$m particle was extremely porous and is likely representative of a common class of particles measured by GIADA on Rosetta. Individual micrometer-sized particles were rare in the MIDAS data set. The COSISCOPE microscope on the dust analyzer instrument COSIMA imaged extremely fragile dust aggregates that fragmented on impact with a target plate at only a few m/s \citep{hornung2016first}. Using laboratory analogs of varying grain sizes and compositions, \citet{rousseau2018laboratory} determined that a mixture of sub-micrometer coal, pyrrhotite, and silicates is a good match for the VIRTIS/Rosetta data. This grain size is consistent with the sizes measured by GIADA, MIDAS and COSIMA on Rosetta. \citet{guttler2019synthesis} categorized C-G dust particles by their structure into three types: 1) a solid group (e.g., a monomineralic fragment), 2) a fluffy group (a dendritic agglomerate of small grains), and 3) a porous group (a more spherical and less-porous agglomerate, similar to fluffy IDPs). The solid and porous particle types are similar to grain morphologies seen in giant cluster interplanetary dust particles.

\begin{figure}[!ht]
\begin{center}
\includegraphics[width=\columnwidth]{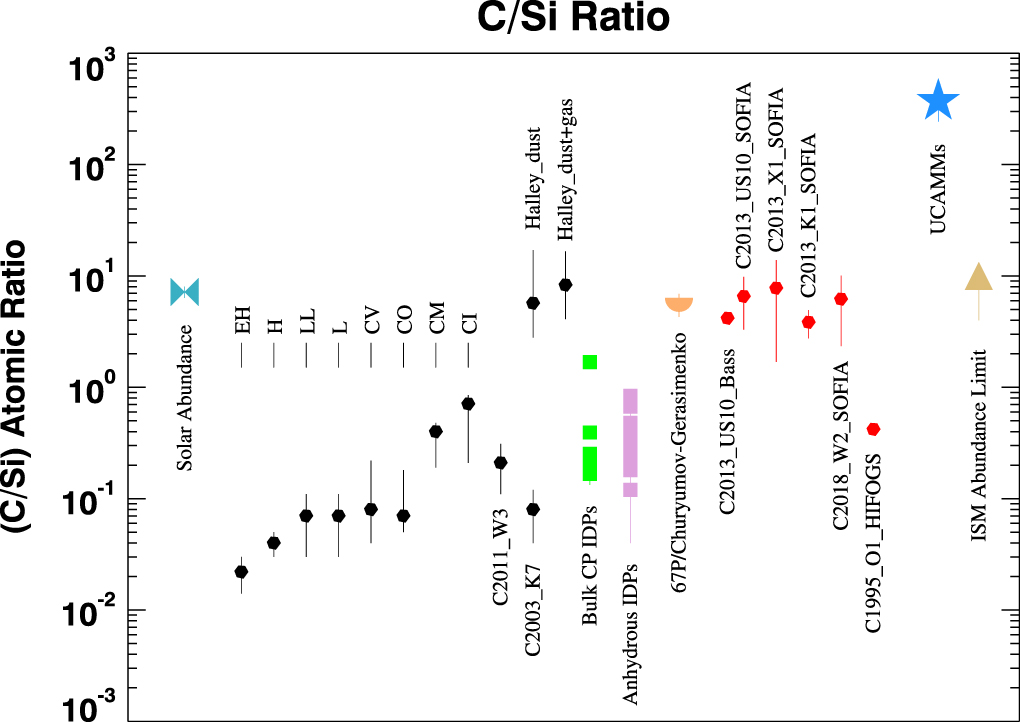}
\end{center}
\caption{Carbon to silicon atom ratios in comets and other objects, from \citet{woodward2021coma}.  
\label{fig:C_to_Si}}
\end{figure}

COSIMA measured the composition of about 250 cometary particles (10~$\mu$m to 1 mm in size). The total C/Si atom ratio of all grains was 5.5$^{+1.4}_{-1.2}$, which is consistent (within uncertainties) with both the Sun and Halley dust, and six times higher than CI chondrites \citep{bardyn2017carbon} (Figure \ref{fig:C_to_Si}). COSIMA measurements of C-G dust showed that it is more similar to carbonaceous chondrites than ordinary chondrites in the major rock-forming elements \citep{stenzel2017similarities}. Sulfur is extremely enriched: a factor of ten higher than the CR chondrite Renazzo \citep{stenzel2017similarities}. It should be noted that COSIMA does not have the maturity of a modern laboratory instrument, and its measurements cannot be independently verified. Additionally, measurements of C-G dust were compromised by PDMS (silicone oil) contamination in the instrument \citep{bardyn2017carbon,stenzel2017similarities}, so care must be taken with interpreting the data.

\subsection{Ultra-carbonaceous Antarctic Micrometeorites, Interplanetary Dust Particles, and Cometary Meteors}
\label{ucammidpmeteors}
The samples in our collection most similar to the carbon-rich Halley dust are ultra-carbonaceous Antarctic micrometeorites (UCAMMs). UCAMMs can have C/Si ratios of $\sim$50 to 1000, which is so high that they may have sequestered carbon out of the gas phase during their formation \citep{dartois2013ultracarbonaceous,dartois2018dome}. Organics in UCAMMs are very nitrogen-rich compared to organics in IDPs and chondrites \citep{dartois2018dome}. \citet{auge2016irradiation} proposed that the N-rich organics in UCAMMs formed via galactic cosmic-ray irradiation of N$_2$-CH$_4$-rich ices on the surfaces of bodies that formed far from the Sun.

Anhydrous and hydrous interplanetary dust particles are collected in the stratosphere in approximately equal abundance \citep{fraundorf1982laboratory}. Most anhydrous IDPs have close to CI elemental composition \citep{schramm1989major} except for moderately volatile elements and carbon, which are more enriched in IDPs by factors of 2--3 compared to CI \citep{flynn1996abundance}. However, individual anhydrous IDPs are highly variable in their carbon content, from 0.6 to 13 times the carbon content of CI chondrites. Some carbon-rich IDPs may be unstudied and hiding in plain sight: they may be mislabeled as terrestrial contaminants in NASA's cosmic dust catalogs (documents that describe dust samples collected in the stratosphere).

Of 20 fragments studied in the large cluster IDP U2-20, five had chondrule-like textures and mineral assemblages \citep{zhang2021oxygen}. Ferromagnesian silicates in anhydrous IDPs are often coated in organic material \citep{flynn2013organic}. IDPs contain intriguing amorphous phases called GEMS (glass with embedded metal and sulfides) which may be the interstellar amorphous silicates observed in the interstellar medium, i.e., the solid building blocks of the Solar System \citep{bradley2022provenance}. However, the origin of GEMS is debated \citep{keller2011origins}. IDPs also contain filamentary enstatite crystals---ribbons, platelets, and whiskers---which likely condensed directly from the gas phase in the solar nebula \citep{bradley1983pyroxene} and may have formed in energetic events in the inner and outer Solar System \citep{utt2023diverse}. GEMS and filamentary enstatite are exceedingly rare in primitive meteorites \citep{villalon2021search}. Stratospheric dust collections can be timed to coincide with a particular meteor shower, with the possibility of collecting dust from a known comet \citep{messenger2002opportunities}. The target comet dust is only a few percent of the total IDP flux, but low concentrations of implanted solar-wind noble gases and lack of solar-flare tracks would be diagnostic of fresh dust from the target comet \citep{messenger2002opportunities,palma2005helium}. Some IDPs collected during the 26P/Grigg-Skjellerup dust stream collection show high (though uncertain, limited by statistics) abundances of circumstellar grains and disordered organic matter with large excesses in D and $^{15}$N (implying a molecular cloud origin). However, the analyzed dust particles were not definitively linked to Grigg-Skjellerup through noble gas analyses or concentrations of solar flare tracks  \citep{busemann2009ultra}.

Meteors from most known cometary meteor have strengths around 1~kPa \citep{trigo2006strength}, within the range estimated for a giant cluster IDP \citep{ogliore2020oxygen} and comet C-G dust \citep{hornung2016first}. This crushing strength is much lower than even the most fragile chondrites (300 kPa for Tagish Lake, \citet{flynn2018physical}). Chemical composition of cometary meteors appears to be similar to IDPs \citep{trigo2003chemical}, though quantitative data sets from cometary meteors are currently limited.

\subsection{Primordial Solar System material}
\label{primordial}
Comets have long been thought to contain the primordial building blocks of the Solar System, dating back to the original nebular hypothesis proposed Immanuel Kant in ``Universal Natural History and Theory of the Heavens'' (1755). The 1981 Walt Disney short documentary ``Comets: Time Capsules of the Solar System'' popularized the idea of comets as vessels of primordial material, a window back into the Solar System's origins. The name ``Stardust'' for the sample-return mission to comet Wild~2 was chosen to evoke this idea. What would we expect such ``primordial'' material to look like?

The Solar System formed in the collapse of a fragment of its parent molecular cloud, which contained dust, organics, ices, and gas. The more refractory components of this mixture, the dust and organics, is the component that could be returned from a non-cryogenic comet sample-return mission like Stardust. \citet{zhukovska2008evolution} uses a model of circumstellar dust (from AGB stars and supernovae), along with dust grown within the molecular cloud, to calculate the relative contributions and origins of silicate, carbon, SiC, and iron dust in the solar neighborhood (Figure \ref{fig:silicatefeature}). The dominant dust sources are silicates and refractory carbonaceous dust grown in the Solar System's parent molecular cloud. This recycled dust is ``isotopically inconspicous''---isotope anomalies in the original circumstellar dust have been averaged away and it would have approximately the same composition as the Sun. The silicates are likely amorphous, as observed astronomically \citep{do2020crystalline}, and possibly similar to GEMS grains found in interplanetary dust \citep{bradley2022provenance}. Recycled carbonaceous dust \citep{jones2014cycling} from our molecular cloud may have $^{12}$C/$^{13}$C close to solar (89) but may contain refractory circumstellar subgrains of anomalous isotopic composition. Graphite grains separated from primitive meteorites \citep{jadhav2013multi} may include recycled refractory carbonaceous dust. 

The C/Si ratio of the interstellar medium is similar to the ratio of the Sun, and much higher than C/Si in CI chondrites (Figure \ref{fig:C_to_Si}). The amount of an element condensed into dust in the ISM is estimated by assuming a gas+dust abundance (usually solar) and then subtracting off the amount observed in the gas phase \citep{bergin2015tracing}. Even when considering only ISM dust (C is more abundant in the gas phase than Si), the C/Si atom ratio is $\sim$6 \citep{bergin2015tracing}, much higher than even the most primitive meteorites. More than 30\% of the carbon in molecular clouds is expected to be in micron-sized carbonaceous dust \citep{herbst2009complex}. Another 30\% of carbon is in simple gas-phase C-bearing molecules like methane, CO, CO$_2$, and methanol (which are frozen out in the coldest clouds). The remaining 20--40\% is in nanometer-sized grains of graphite and hydrogenated amorphous carbon \citep{draine2003interstellar}, in polycyclic aromatic hydrocarbons (PAHs) \citep{tielens2008interstellar}, and in other complex organic molecules. 

More than 50 complex organic molecules (C-bearing species with more than six atoms) are detected in the interstellar and circumstellar media, nearly all in the gas phase \citep{herbst2009complex}. It is likely that complex organic molecules can be formed on ice mantles of interstellar grains through, e.g., energetic particle irradiation \citep{chuang2021formation}.  Even the most primitive meteorites have undergone alteration on their parent bodies which changes the nature of their organics \citep{alexander2017nature}, and have a bulk C/Si ratio much lower than the molecular cloud solids (Figure \ref{fig:C_to_Si}). Primordial Solar System solids should have high C/Si and contain a suite of pristine interstellar organic molecules, including phases not detected astronomically due to their low abundance in the gas phase. Organic molecules can be detected in bright comets such as Hale-Bopp \citep{bockelee2004composition}, but a sample of primordial Solar System material would allow for the detection and characterization of the entire suite of organics with sensitive lab techniques.

\section{Comet Wild 2 and the Stardust Mission}
\label{wild2andmission}

For nearly its entire lifetime, comet 81P/Wild 2 was in an outer Solar System orbit where temperatures did not exceed 50 K. On September 9\textsuperscript{th}, 1974, the comet passed 0.0061 AU from Jupiter (about 13 Jupiter radii), which changed its orbital period from 43 years to 6.2 years, its aphelion from 25 AU to 5.24 AU, and its perihelion from 4.97 AU to 1.59 AU \citep{sekanina2003model}.

Observations of comet Wild~2 show that abundant volatile gases are being emitted from its coma, such as C$_2$H$_2$, C$_2$H$_6$, HCN and NH$_3$ \citep{russo2014volatile}. It does not appear to be depleted in volatiles compared to other Jupiter-family comets (though each JFC is chemically distinct in its volatile composition). It is optically characterized as carbon-chain depleted \citep{a1995ensemble,fink2009taxonomic} but its 
volatile abundances (including C$_2$-bearing species relative to H$_2$O) appear to be more similar to carbon-chain-typical JFCs such as 2P/Encke, 6P/d'Arrest, 10P/Tempel 2, and 103P/Hartley 2. However, Wild~2 has higher C$_2$H$_2$ and C$_2$H$_6$ abundances than these other JFCs \citep{russo2014volatile}.

The Stardust mission was selected in 1995 as a NASA Discovery mission. The primary objectives of the mission were to collect cometary dust from comet 81P/Wild 2 (a Jupiter-family comet) and interstellar dust in silica aerogel \citep{tsou2003wild} and return the samples to Earth for analysis. Stardust was launched aboard a Delta II rocket from Cape Canaveral on February 7 1999. Candidate interstellar dust grains were identified in the interstellar collector \citep{westphal2014evidence}, but will not be discussed further here.

\begin{figure}[!ht]
\begin{center}
\includegraphics[width=0.9\columnwidth]{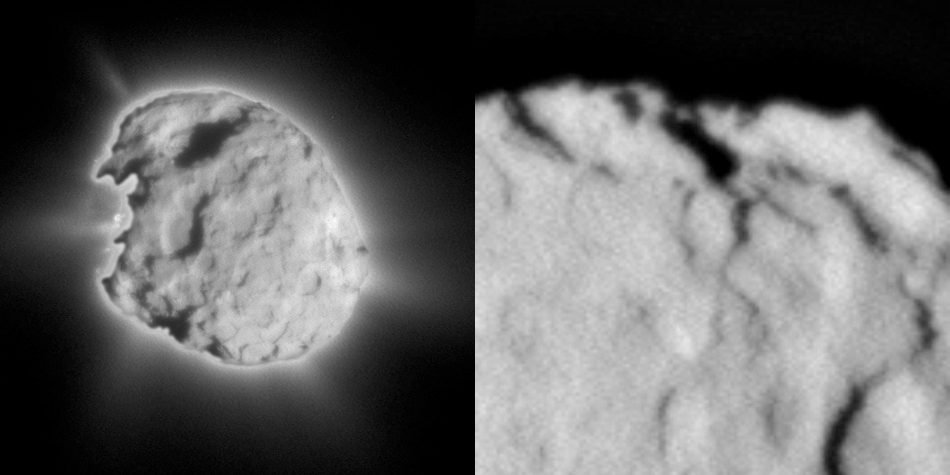}
\end{center}
\vspace*{-0.25in}
\caption{Left: Composite image of a long-exposure and short-exposure image of the nucleus of comet Wild 2, showing surface features and jet activity. Right: Close-up view from a different image showing pinnacles, depressions, and ridges on Wild 2. \label{Wild2nucleus}}
\end{figure}

The Stardust spacecraft came within 234 km of Wild 2's 5-km nucleus on January 2\textsuperscript{nd}, 2004 when the comet was 1.86 AU from the Sun. Wild 2's nucleus was imaged by the spacecraft's optical navigation camera with the closest image recording $\sim$14 m per pixel (Figure \ref{Wild2nucleus}). The images recorded multiple jets being emitted from the comet's surface. The surface of the nucleus is topographically complex, with surface mesas, surface depressions (some larger than 0.5 km), and pinnacles \citep{bro04}.  

\subsection{Capture of Wild 2 particles}
The Stardust spacecraft flew through the coma of comet Wild 2 with a relative speed of 6.1 km/s and collected thousands of particles in a 0.1 m$^2$ collector of low-density (0.01--0.05 g/cm$^3$) silica aerogel (85\% surface area) tiles supported by an aluminum-foil-covered frame (15\% surface area). Cometary particles impacting the aluminum foil produced craters and partial melting of the impactor; impacts into aerogel produced tracks with variable amounts of thermal modification of the captured fragments.


\subsubsection{Aerogel}
Aerogel tracks in the Stardust collector vary in shape from long and skinny ``carrot'', or Type A, tracks to short and wide ``bulb'' tracks, with short (Type B) or long (Type C) styli created by larger terminal particles. Analog test shots fired into aerogel have shown that the shape of the track is determined largely by the subgrain size and bulk strength of the impactor (rather than its density). Monomineralic grains produce Type A tracks, and powdered impactors produce Type B and C tracks \citep{kea12}. 

Stardust aerogel tracks are predominantly type A (65\%) and type B (33\%) with a small fraction (2\%) type C. A large fraction, $\sim$30\%, of the Type A tracks could be classified as A* (small, squat tracks)  \citep{kea12} that could be made by volatile-rich impactors. However, most of the Stardust tracks are either carrot-shaped or have a significant stylus. This observation alone indicates that most of the Wild 2 impactors contained either large monomineralic grains or large rocky fragments with high internal strength. Tracks with large bulbs likely contained friable or volatile-rich material, much of which ended up embedded in the walls of the track's bulb. Cometary material in the track bulb is usually partially melted and mixed with molten-solidified aerogel, making it difficult to analyze.

\begin{figure}[!ht]
\begin{center}
\includegraphics[height=0.25\textheight]{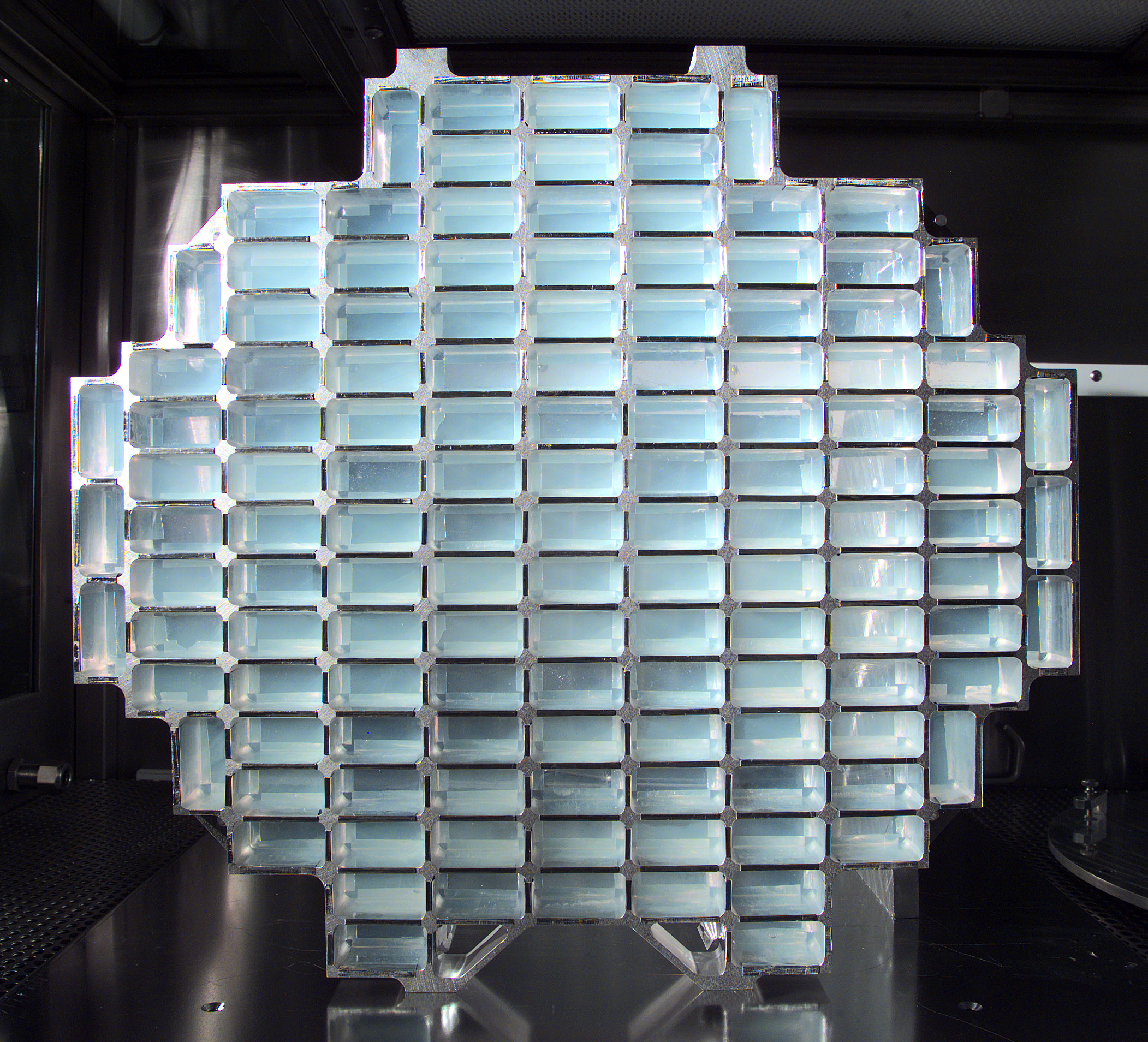}
\includegraphics[height=0.25\textheight]{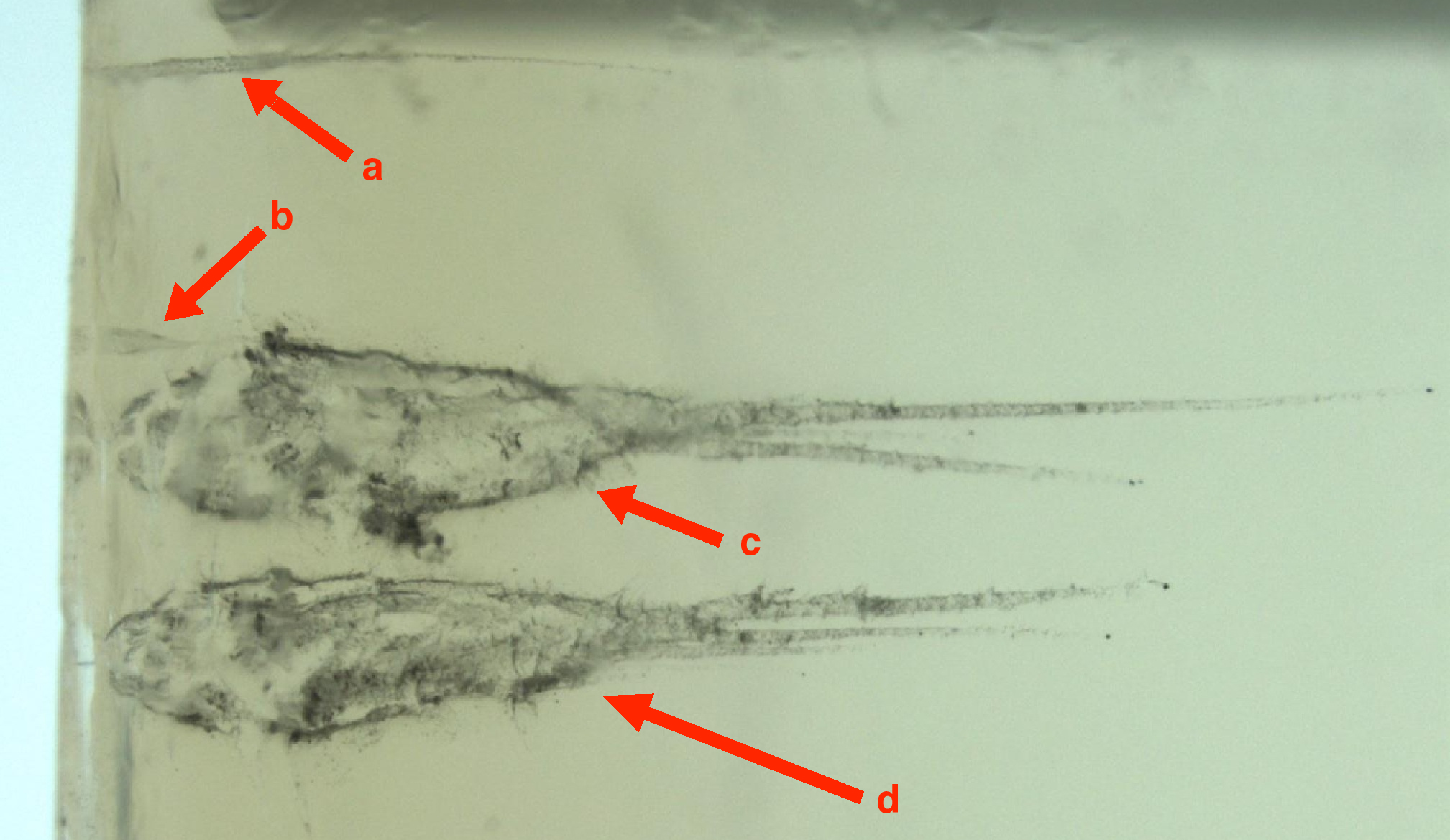}
\end{center}
\vspace*{-0.25in}
\caption{Left) Stardust collector. Right) Four cometary tracks in aerogel tile C2027: two large Type C Stardust aerogel tracks (c \& d: C2027,6: 8.1mm, C2027,7: 8.5mm), a very small Type C (b: C2027,8: 1.5mm) and a Type A whisker (a: C2027,9: 4.7mm). \label{tracks}}
\end{figure}

Heating due to capture of cometary particles at 6.1 km/s caused severe modifications to the smallest (sub-$\mu$m) component of captured Wild 2 material, but only minor modification to particles larger than $\sim$2~$\mu$m. Well-preserved fine-grained material can be found on the trailing end of larger grains \citep{stodolna2014characterization}. Unprotected sub-$\mu$m grains were in close proximity to molten aerogel at nearly 2000$^{\circ}$ C and were often melted, though some can be found that are better preserved \citep{leroux2012fine}. Larger silicate grains in the aerogel are usually very well preserved, including Wark-Lovering rims on refractory particles \citep{joswiak2017refractory}. This preservation is due to the formation of a protective shell during aerogel capture, and was seen in 6~km/s light gas-gun shots of soda lime glass spherules before launch of the Stardust mission \citep{horz1998capture}. The surface of the glass (melting point $\sim$1000$^{\circ}$ C) did not melt, even though it was surrounded by compressed unmelted aerogel with an outer layer of silica melt ($\sim$2000$^{\circ}$ C).

\subsubsection{Al Foils}
Aluminum foils covered the support structure of the collector frame (Figure \ref{tracks}) and provided an additional collection medium. The foils could be removed from the frame, flattened, and analyzed in an SEM to search for impact craters containing residue. Cometary material in the impact craters had been subjected to much greater heating and alteration than aerogel-captured particles, but it is more concentrated and the foils are easier to analyze by some analytical techniques. However, like the aerogel, the foils contain impurities that can complicate analyses. 

Analyses of residue in Al foil impact craters show that some mineral grains can be surprisingly well preserved and the mineralogy of sub-$\mu$m grains can be determined \citep{leroux2008transmission}. Even less refractory phases, like sulfides, can be preserved in the impact residues \citep{haas2020fib}.


\subsection{Limitations of the Stardust comet Wild 2 samples}
\label{statanec}

The amount of material that the Stardust mission returned to Earth is very small ($\sim$1~mg) compared to other samples of Solar System materials, such as carbonaceous chondrites and lunar rocks returned by the Apollo missions. Additionally, the samples are extremely challenging and time consuming (and thus, expensive) to prepare and analyze. Only a few groups in the world have developed the techniques to reliably prepare and measure Wild~2 samples. Therefore, it is challenging to make meaningful statistical assessments of components of the Wild 2 samples. For example, a statistically significant measurement of the abundance of type II chondrules in comet Wild 2 compared to unequilibrated ordinary chondrites would require analyses of dozens of tracks over many years. Such analyses are extremely valuable, and are continuing in a number of laboratories around the world. However it is also possible to learn a lot about the comet through anecdotal observations. The definitive presence of a CAI fragment in the comet requires a process to get this object into the comet nucleus, but the efficiency of this mechanism can only be determined by a statistical assessment of the abundance of CAIs in Wild 2 compared to, e.g., CR chondrites.

Aerogel capture and the fine-grained nature of the collected material creates additional challenges. The Wild~2 samples have been abraded during capture, and often their shapes and sizes are not the same as they likely were in the comet nucleus. The shapes of objects in meteorites are critical clues to their origin. For example, particles with mineralogy/petrology similar to chondrules in the Wild~2 samples are usually called ``chondrule fragments'' or ``chondrule-like'' because their shapes have been ablated during capture, and the round shape of a chondrule testifies to its origin as a free-floating melt droplet in space.  Associations between neighboring mineral grains can also be lost during capture: two adjacent grains can end up in different places in the aerogel track. The small size of Wild~2 grains makes it difficult to compare to analogous objects in meteorites with much larger grain sizes. 

Finally, the Stardust mission sampled only a single comet. Spectroscopic observations of comets tells us that comets are enormously diverse in their volatile content \citep{biver2015chemical}, Si/C ratio \citep{woodward2021coma}, and silicate composition \citep{kelley2009composition} (though the inferred Mg/Fe ratio of silicates from remote sensing is complicated by particle-size, fragmentation, and other effects, and Wild~2 showed a wide range of silicate compositions in a single comet). The diversity of solid material in comets can be better explored through analyses of stratospheric IDPs, which most likely come from a large number of comets \citep{nes10}.


\section{Identified Phases in the comet Wild 2 Sample}
\label{phasesidentified}

The most common mineral phases in the Stardust samples are olivines, pyroxenes, and sulfides of varying compositions and formation histories. Other phases identifed include Fe-Ni metal (kamacite and taenite), phosphates (merrillite and whitlockite), phosphides (e.g., schreibersite and barringerite), silica (tridymite), iron-oxides (magnetite, hematite), nitrides and carbides (osbornite and silicon-carbide), and Pt-group metals. These phases are also found in chondrites, but their composition, abundance, assemblages, and distributions in Wild~2 are often distinct from chondrites, as discussed below.

Native cometary amorphous phases in the Stardust samples are difficult to identify due to the effects of hypervelocity capture in aerogel \citep{ishii2008comparison}. \cite{wes09} report a one-sided 2$\sigma$ lower limit on the crystalline fraction of Fe-bearing Wild 2 grains of 50\%, based on synchrotron Fe K-edge XANES measurements of 194 fragments in 11 tracks. \citet{stodolna2012mineralogy} found that $\sim$50\% of cometary material in the wall of Track C2092,3,80 (a Type B track) is crystalline. Comet Wild~2 is not made mostly of amorphous materials, but the precise abundance and nature of its amorphous phases remains somewhat obscured by the hypervelocity capture process.

Organics from comet Wild~2 have also been found in the Stardust samples, though aerogel contamination and the effects of hypervelocity capture can make studies of Wild~2 organics challenging, as discussed in Section \ref{organics}.

Notable phases that have not been conclusively identified in the Stardust samples are phyllosilicates and carbonates, which are abundant in aqueously altered chondrites. Relict phyllosilicates were reported by \citet{schmitz2011relict}, though no other studies found evidence for phyllosilicates in the Stardust samples.  Carbonates were reported by \citet{wirick2007carbonates}, \citet{mikouchi2007mineralogy}, and others, though these phases (especially calcium carbonates) may be aerogel impurities (calcium is a common aerogel contaminant). Based on spacecraft mass-spectrometery measurements of comet Halley dust, refractory carbon would be expected to be a major component of the larger Stardust particles. In most chondrites, carbon is contained in fine-grained material as, e.g., organic matter. Refractory carbon is nearly absent in the larger collected particles, even though it should survive aerogel capture \citep{zubko2012evaluating}. This implies that the Wild~2 carbon abundance is more similar to carbonaceous chondrites and less similar to Halley dust. If carbon is present in Wild~2 at the percent-level, it is likely in the fine-grained $<2$~$\mu$m size fraction.

\section{Igneous Rocks: CAI, AOA, and chondrule fragments}
\label{chondruleaoacai}
Igneous rocks are the dominant component of the larger ($> 2~\mu m$) particles collected from comet Wild~2 \citep{bro12}. These particles have similar mineralogy as chondritic meteorites, though the shapes of the Wild 2 objects are not well preserved, and often the size scales are different. 

\subsection{CAI fragments}
\label{caifragments}
One of the first tracks analyzed after the samples returned to Earth in 2006 was Track 25 (C2054,4,25,0,0). This track contained fragments with mineralogy similar to calcium-aluminum inclusions (CAIs) found in chondrites \citep{simon2008refractory}. Calcium-aluminum-inclusion are the oldest Solar System objects found in the meteorite record, with an age of 4567.30 Ma \citep{connelly2012absolute}. The texture, mineralogy, and isotopic composition of CAIs indicate they formed from a combination of condensation, aggregation, melting, and evaporation in a hot ($>$1300~K) solar-like gas near the proto-Sun \citep{krot2019refractory}. The fragment Inti from Track 25 is $\sim$30 $\mu$m in size, and made up primarily of anorthite, fassaite, Al-diopside, spinel, and melilite, with small grains of osbornite (TiN) and perovskite. A total of 23 CAI fragments, including Inti, have been found in Track 25 and are almost certainly all related to each other \citep{joswiak2017refractory}. Inti's oxygen isotopic composition is $^{16}$O-rich ($\delta^{17,18}$O$=-40$\permil, where $\delta^{17,18}$O is the parts-per-thousand deviation from a standard $^{17,18}$O/$^{16}$O ratio), similar to many other meteoritic CAIs \citep{simon2008refractory}. Additionally, Inti is very similar in texture and mineralogy to a fine-grained CAI found in the CV chondrite Leoville \citep{krot2004fine}. Orbornite is very rare in meteorites, but has been found enclosed in grossite within CAIs in the CH/CB chondrite Isheyevo \citep{meibom2007nitrogen}, in an osbornite-spinel clast in the CH chondrite ALH 85085 \citep{weisberg1988petrology}, and in chondrules in an enstatite chondrite \citep{buseck1972mineralogy}.

Three other CAI-like fragments have been identifed in the Stardust samples. Coki (from track 141, C2061,3,141,0,0) is a $\sim$5~$\mu$m particle that consists mostly of anorthite with spinel inclusions, and Al-Ti diopside \citep{matzel2010coki}. Coki is made of minerals that are less refractory than Inti and formed with initial $^{26}$Al well below the canonical value (see Section \ref{26al}). WF216 is a 2-$\mu$m nodule from the bulb of track 172 (C2119,1,172,0,0) that is mineralogically zoned: the core is spinel, then anorthite, then pyroxene. The spinel and anorthite contain very small ($<$10~nm) osbornites \citep{joswiak2017refractory}. TuleF4 is a $\sim$1~$\mu$m CAI nodule from track 80 (C2092,2,80,0,0). It consists of spinels poikilitically enclosed in anorthite, all enclosed in a pyroxene shell \citep{joswiak2017refractory}. 

CAI fragments make up approximately 0.5 vol\% of Wild 2 material \citep{joswiak2017refractory}, which is comparable to some carbonaceous chondrite groups but higher than enstatite and ordinary chondrites. Refractory particles that are likely CAI fragments have been identified in interplanetary dust particles \citep{mckeegan1987oxygen,zolensky1987reports,joswiak2017refractory}. The abundance of CAI fragments in the giant cluster interplanetary dust particle U2-20 was estimated to be $\sim$1 vol. \% \citep{joswiak2017refractory}. For comparison, CAIs are $\sim$1\% in CO and CM chondrites, 3\% in CV chondrites, and 0.01-0.2\% in H chondrites \citep{rus98}. CAIs found in the comet Wild 2 and U2-20 samples studied by \citet{joswiak2017refractory} are of moderately refractory nature. The highest-temperature minerals---grossite, hibonite, and corundum---are not found in either the comet Wild 2 or U2-20 samples, though it is important to keep in mind the limited number of Wild~2 analyses compared to chondrites (Section \ref{statanec}). However, hibonite has been found in three refractory IDPs reported by \citet{zolensky1987reports}.

\subsection{Amoeboid Olivine Aggregates}
\label{aoa}
Amoeboid olivine aggregates (AOAs) in chondrites consist of forsterite, Fe-Ni metal, and Ca- and Al-rich minerals also found in CAIs (spinel, diopside, and anorthite). Unaltered AOAs are uniformly $^{16}$O-rich: $\delta^{17,18}$O $\approx -40$\permil\ with little variation \citep{fagan2004oxygen,ushikubo2017long}. \citet{fukuda2021correlated} studied three Wild~2 $^{16}$O-rich forsterites (T57/F10, T77/F50, and T175/F1) and AOAs in the primitive CO chondrite DOM 08006 by measuring trace-element concentrations (Cr, Ca, Mn, Fe), O, and Mg isotopes by SIMS. AOAs in the CO chondrite have negative $\delta^{25,26}$Mg values and specific correlations between Cr, Ca, Mn, and Fe. One Wild~2 forsterite, T57/F10 (from Track 57, C2009,2,57,0,0), shows similar Mg isotopes and trace-element concentrations as a porous CO-chondrite AOA, and so likely has a similar origin. The other two Wild~2 particles did not have isotopic and trace-element concentrations similar to the CO-chondrite AOAs. Two other reports of $^{16}$O-rich forsterites in the Stardust samples may be related to AOAs \citep{nakamura2011nanometer,Nakashima2012}, but more detailed analyses of Mg isotopes and trace-element concentrations are needed.

\subsection{Chondrule fragments}

Aluminum-rich chondrules have compositions intermediate between ferromagnesian chondrules and CAIs \citep{krot2002anorthite}, but have faster cooling rates and higher peak temperatures than CAIs that are more consistent with ferromagnesian chondrules \citep{tronche2007formation}. Two probable Wild~2 Al-rich chondrules have been reported. A 4$\times$6~$\mu$m terminal particle in track 130 (C2061,3,130), ``Bidi''. is made of Fo$_{97}$ olivine, An$_{97}$ feldspar, and Al-, Ti-bearing clinopyroxene with augite/pigeonite lamellae that indicate fast cooling \citep{joswiak2014terminal}. Based on its mineralogy, chemistry, and $^{16}$O-poor composition, \citet{joswiak2014terminal} conclude that Bidi is most similar to an Al-rich chondrule. \citet{bridges2012chondrule} report a 6$\times$2~$\mu$m particle in track 154 (C2063,1,154,1,0) made of Al-rich diopside and pigeonite with accessory forsterite and enstatite. They conclude that the Al-rich/Ti-poor composition and O isotope composition is most similar to Al-rich chondrules in carbonaceous chondrites.

Chondrules are sub-millimeter-sized spherules that have undergone significant melting and compose  15--75 vol.\ \% of chondrites (except CI chondrites, though they likely contain chondrule fragments) \citep{rubin2000petrologic}. Chondrules can be divided into type I (Mg\# $>$ 0.9) and type II (Mg\# $<$ 0.9) with the latter forming in more oxidizing conditions \citep{villeneuve2015relationships}. Lone ferromagnesian fragments found in chondrite matrices appear to be related to chondrules \citep{jacquet2021origin}, and the same may be true for lone mineral fragments found in Stardust tracks. Many particles are multi-phase and have diagnostic chondrule features like triple junctions, porphyritic or poikilitic textures, glassy mesostasis in contact with silicates, and rounded Fe metal inclusions \citep{nak08,gainsforth2015constraints}. Type II ferromagnesian chondrule fragments (see Section \ref{statanec}), like Iris  \citep{gainsforth2015constraints} (Figure \ref{chondruleaoacai}B), are abundant in the comet Wild 2 samples. Type I chondrule fragments, like Gozen-sama \citep{nak08} (Figure \ref{chondruleaoacai}A), are much less common than type II fragments. Other chondrule fragments are described in, for example, \citet{bridges2012chondrule}, \citet{joswiak2012comprehensive}, and \citet{jacob2009pyroxenes}. The large (17-mm long) Track 227 (C2014,3,227,0,0) contains 13 chondrule-like fragments, including a mostly intact 60$\times$50~$\mu$m type II chondrule (T227,8,F1) with porphyritic texture. This spectacular cometary chondrule is composed of Fo$_{62}$ olivine, Fe-Ni metal, chromite, and augite in an albitic glassy mesostasis \citep{zhang2023rocky}. An important property of Wild~2 chondrule fragments is that that there are no observations of alteration of mesostasis from contact with liquid water. This distinguishes them from even the lowest petrographic grade UOC and CC chondrules \citep[e.g.,][]{grossman2002zoned,noguchi2021mineralogy}.

\begin{figure}[!ht]
\begin{center}
\includegraphics[width=\columnwidth]{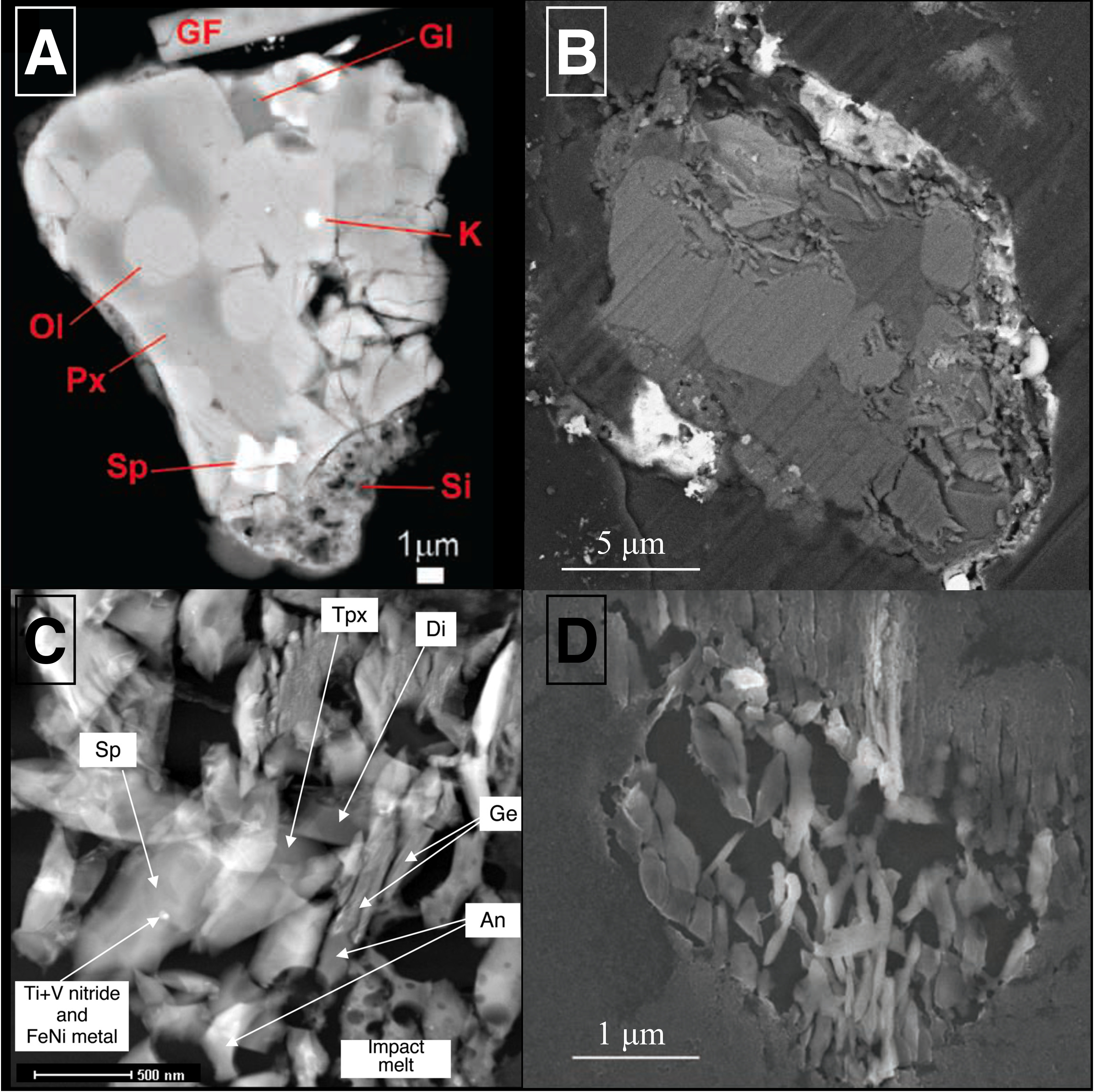}
\end{center}
\vspace*{-0.25in}
\caption{Examples of chondrule and CAI fragments found in the Stardust cometary samples. A) BSE image of chondrule fragment ``Torajiro'' \citep{nak08}; B) BSE image of Type II chondrule fragment ``Iris'' \citep{ogl12}; C) HAADF TEM image of CAI fragment ``Inti'' \citep{sim08} D) SE image of Type C CAI fragment ``Coki'' \citep{mat10}.  \label{chondrulescais}}
\end{figure}

\subsubsection{Major and minor element concentrations in ferromagnesian silicates}
\label{minorelementsinsilicates}
\begin{figure}[!ht]
\begin{center}
\includegraphics[width=0.75\columnwidth]{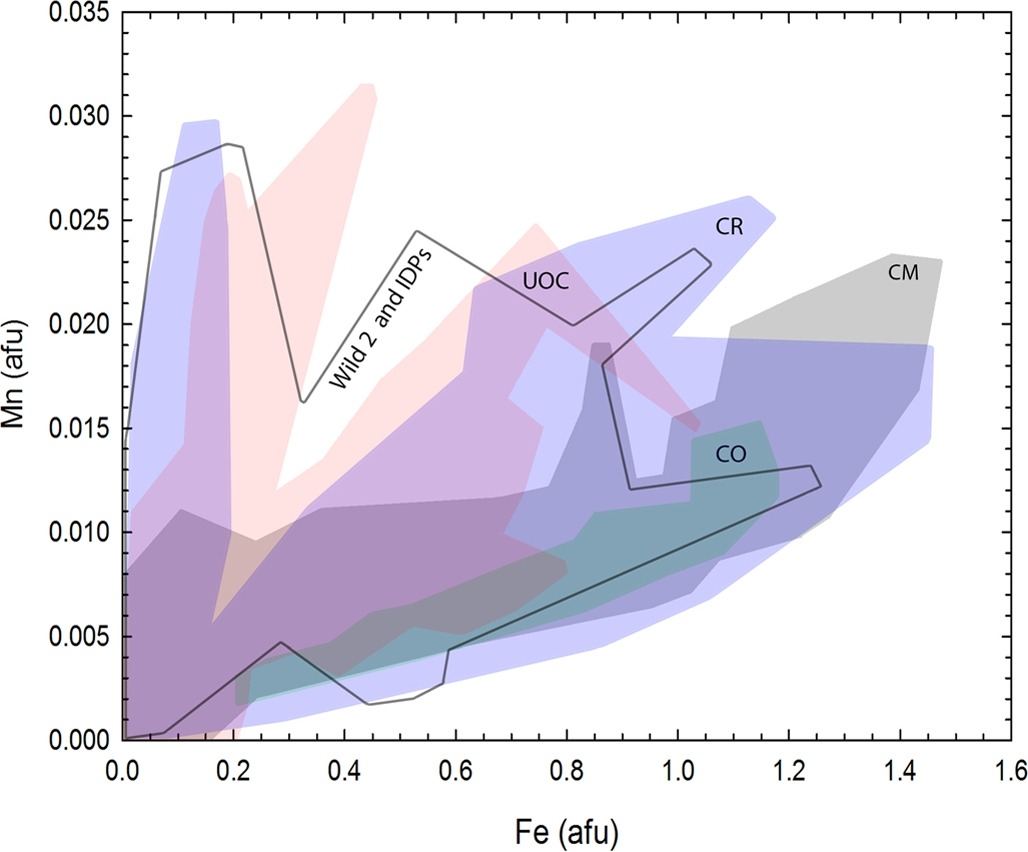}
\end{center}
\caption{Mn vs.\ Fe in Wild~2 olivine compared to meteorite groups (UOC, CR, CM, CO) and IDPs, from \citet{schrader2022prolonged}. Units are atoms per formula unit. \label{Schrader_olivine}}
\end{figure}

Major and minor element concentrations in Wild~2 ferromagnesian silicates can be compared with chondrite samples to determine the comet's possible asteroidal similarities. In Wild~2, the Fe/(Fe+Mg) ratio in olivine shows a very broad distribution compared to matrix grains in different groups of chondrites \citep{frank2014olivine}. Each chondrite group has chondrules with characteristic Fe-Mn compositions which are controlled by the formation conditions of the chondrules. Meteorite groups accrete ``local'' materials that formed under similar conditions, so it is not surprising that chondrule olivine in a given meteorite group has a well-defined Mn/Fe ratio. The Mn/Fe ratio in Wild~2 olivine is highly variable and nearly spans that of olivines measured in all chondrite groups  \citep{frank2014olivine,brownlee2017diversity} (Figure \ref{Schrader_olivine}). This diversity of compositions implies Wild~2 accreted silicates from disparate regions of the Solar System, including material sampled by OC, CO, and CR type II chondrules \citep{brownlee2017diversity}. The spread of Fe-Mn compositions in CR chondrules is similar to Wild~2 (Figure \ref{Schrader_olivine}) \citep{brownlee2017diversity}. Some Fe-Mn compositions of Wild~2 olivines are found only in pristine ordinary chondrites \citep{schrader2022prolonged}. The authors conclude that this requires inner Solar System chondrules to be transported outward beyond Neptune for incorporation into Wild~2. 


\subsubsection{Oxygen isotopes: $^{16}$O-rich grains}
\label{oisotopes_o16rich}
Relict grains in chondrules are silicate grains (usually olivine) that survived the chondrule formation process without melting \citep{rambaldi1981relict}. They did not equilibrate with their host chondrule and have distinct chemical and isotopic compositions. Relict grains may be related to AOAs \citep{marrocchi2018oxygen}, early-formed planetesimals \citep{libourel2007evidence}, or a previous generation of chondrules \citep{ruzicka2007relict}. \citet{ushikubo2017long} report $^{16}$O-rich relict grains in the carbonaceous chondrite Acfer 094 which likely are genetically linked to AOAs or accretionary rims of CAIs. Similar $^{16}$O-rich relict grains have been found in other chondrites \citep[e.g.,][]{schrader2013formation,tenner2013oxygen}. Gozen-sama (C2081,1,108,1) is a type I chondrule fragment from comet Wild~2 with an $^{16}$O-rich relict olivine: $\delta^{17}$O=$-$47\permil\ , $\delta^{18}$O=$-$50\permil\ \citep{nak08}. \citet{zhang2023rocky} report $^{16}$O-rich relict grains in track 227, likely associated with the large type II chondrule terminal particle described above. Relict olivines in chondrule-like objects studied by \citet{zhang2023rocky} and \citet{nak08} contain relict olivine that has equilibrated in Fe but not oxygen isotopes. Fe has diffused in each particle to a uniform Mg\# but the $^{16}$O-rich composition is retained because of the slower diffusion coefficient of oxygen \citep{ruzicka2007relict}.

The relative abundance of $^{16}$O-rich to $^{16}$O-poor phases in comet Wild~2 is higher than other carbonaceous chondrite groups except for CI \citep{kawasaki2022oxygen}. Samples returned from asteroid Ryugu are very similar to CI chondrites \citep{ito2022pristine} and also have a high fraction of $^{16}$O-rich phases, similar to comet Wild~2 \citep{kawasaki2022oxygen}. In Wild~2, pyroxenes and olivines enriched in $^{16}$O \citep{defouilloy2017origin,fukuda2021correlated} may be related to the Kakangari meteorite (see Section \ref{reducedphases}). CAI fragments, found in similar abundance in Wild~2 as in carbonaceous chondrites \citep{joswiak2017refractory}, tend to also be $^{16}$O-rich. 

The high abundance of $^{16}$O-rich grains in comet Wild~2 and Ryugu may be due to the efficient outward transport of early-formed Solar System solids with O isotope composition similar to the Sun. Grains that condensed early ($<$0.5 Myr after CAIs) and close to the Sun would be efficiently transported outwards because the rate of outward transport is highest when the mass accretion rate onto the proto-Sun is highest \citep{armitage2001episodic,boss2020evolution,ciesla2010distributions}. 

\subsubsection{Oxygen isotopes: $^{16}$O-poor grains}
\label{oisotopes_o16poor}
\begin{figure}[!ht]
\begin{center}
\includegraphics[width=0.75\columnwidth]{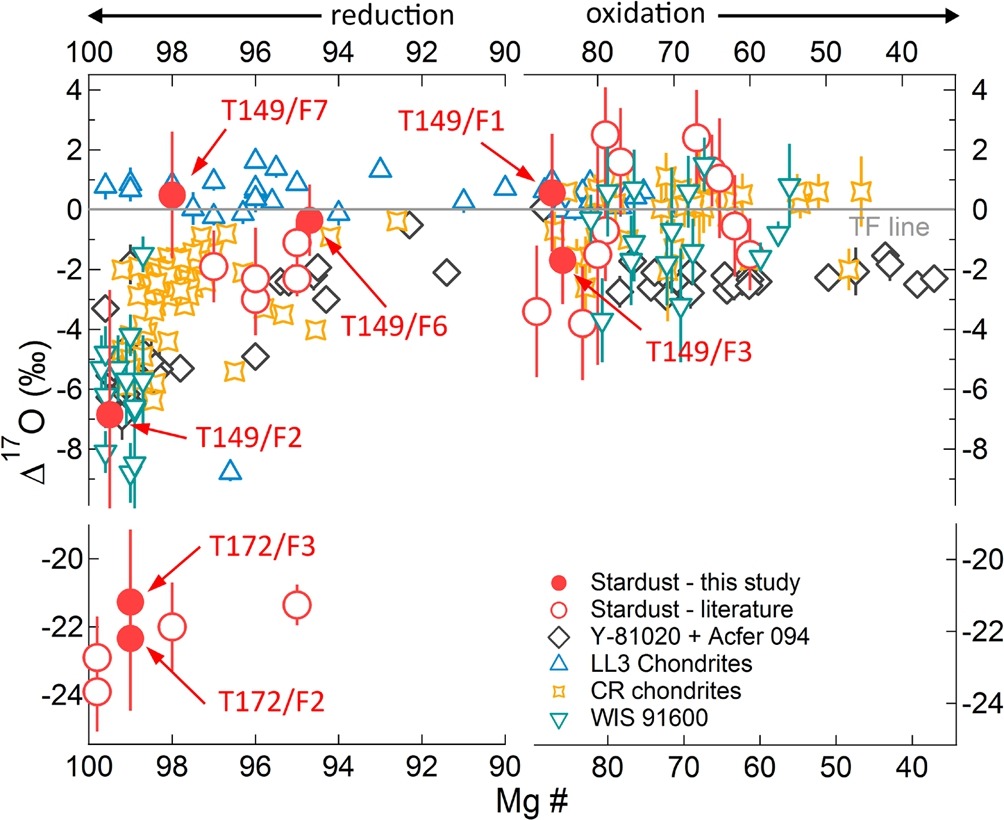}
\end{center}
\caption{$\Delta^{17}$O vs.\ Mg\# for Wild~2 silicates \citep{nak08,nakashima2012oxygen,ogliore2012incorporation,ogliore2015oxygen,defouilloy2017origin} compared to primitive chondrite chondrules. WIS 91600 is a Tagish Lake-like chondrite. Figure is from \citet{defouilloy2017origin}. \label{defouilloy_O_isotopes}}
\end{figure}

High-precision triple oxygen isotope compositions, measured by large-geometry SIMS techniques have been reported for 37 comet Wild~2 grains \citep{nak08,nakashima2012oxygen,ogliore2012incorporation,ogliore2015oxygen,defouilloy2017origin,zhang2022intra}. Variation in $\Delta^{17}$O (the actual $\delta^{17}$O value minus the predicted $\delta^{17}$O value if the O isotope ratios followed the terrestrial mass-dependent fractionation law) is from about $-7$\permil\ to $+2$\permil, indicating that these particles formed in different Solar System reservoirs. The distribution in $\Delta^{17}$O is very close to that of CR chondrules, and the $\Delta^{17}$O vs.\ Mg\# trend (Figure \ref{defouilloy_O_isotopes}) is also very similar \citep{defouilloy2017origin}. However, $^{16}$O-rich relict grains which are found in Wild~2 are very rare in CR chondrules \citep{zhang2023rocky,tenner2015oxygen}. Preliminary results from \citet{zhang2023rocky} (not included in the 37 grains discussed above) describe very FeO-rich silicates in track 227 with $\Delta^{17}$O up to $+7$\permil\ that are not seen in CR, but are found in CH-CB chondrites. \citet{zhang2021oxygen} measured O isotopes in IDPs and find a similar trend in $\Delta^{17}$O vs.\ Mg\# as seen in comet Wild~2.

\section{Presolar Grain Abundance}
\label{psg}
One of the highest priority goals of the Stardust mission was to determine the circumstellar or ``stardust'' grain abundance of comet Wild 2. If the rocky material of comets was sourced from the pristine ingredients that formed the Solar System, isotopically anomalous ancient circumstellar grains could make up a large fraction of the solid material of Wild 2. The matrices of primitive chondrites contain small ($<$1~$\mu$m) isotopically anomalous circumstellar silicates at the $\sim$200~ppm level \citep{floss2016presolar}. However, since even the most unaltered meteorites experienced parent-body heating and/or aqueous alteration, the abundance of circumstellar grains in the solar nebula may have been higher.  Such destructive processes were most likely insignificant on comet Wild 2 (see Section \ref{aqueous}), so presolar grain abundances were speculated to be many times higher in Wild~2 than that seen in even the most primitive meteorites.

The first 18 years of isotopic analyses of Wild 2 grains has disproved the hypothesis that a substantial fraction of the solid material of this comet is made of isotopically anomalous circumstellar grains. This is a surprising and unexpected result. Only six presolar grains have been identified: five oxides or silicates 100--300 nm in diameter with anomalous O isotopic composition found in the foils \citep{flo13}, and one 400~nm SiC grain with anomalous C isotopes found in an aerogel track \citep{messenger2009discovery}.

High-speed impact of the cometary material into Al foil can destroy presolar grains. To account for this effect, \citet{flo13} fired test shots of the Acfer 094 chondrite into Al foil at Stardust-encounter speeds ($\sim$6 km/s) and measured the presolar grain abundance in the crater residue. They found that the impact residue was depleted in presolar grains compared to measurements of unshot Acfer 094, and concluded that the same loss/destruction processes were present in the Stardust foil samples. With this correction factor, \citet{flo13} calculated the silicate+oxide presolar grain abundance in comet Wild 2 to be 600--830 ppm. The confidence interval represents only the range of Acfer 094 presolar grain abundance measurements.

However, an estimate of the presolar grain abundance of Wild 2 is a statistical estimate and the true confidence interval must take statistical uncertainties into account. Assuming the simplest scenario where presolar grains are Poisson distributed in the Acfer 094 test shots and Stardust samples, the 1$\sigma$ confidence interval from the \citet{flo13} measurements is $\sim$300--3700 ppm and the 2$\sigma$ lower bound is $\sim$140 ppm.

With these uncertainties, it is not possible to confidently distinguish the O-rich presolar grain abundance in Wild 2 from the matrix of primitive meteorites \citep{flo09} or IDPs \citep{flo06}. The current measurements still allow for comet Wild 2 to be ten times richer in presolar grains than the most presolar-grain rich Solar System materials. Acquiring a statistical estimate of rare grains in a microscopic sample is a very challenging problem. 

\section{Aqueous alteration and thermal metamorphism}
\label{aqueous}

The CI, CM, and CR carbonaceous chondrites contain a variety of secondary minerals that were formed by aqueous alteration \citep{zol88}. Unequilibrated ordinary chondrites show evidence for less-pervasive aqueous alteration \citep{kro97}, as do CV carbonaceous chondrites \citep{bre97}.  The aqueous alteration that these meteorites experienced most likely occurred on the asteroidal parent body, though some types of aqueous alteration could have occurred in the solar nebula \citep{bre03}.  The decay of $^{26}$Al and impacts can generate heat in an asteroid to melt ice and facilitate aqueous alteration. 

Comet Wild 2, however, is thought to have remained in cryogenic conditions for its entire life before its close encounter with Jupiter in September 1974. Heating by the Sun probably hasn't significantly affected Wild 2's sub-surface material as prior to 1974 the comet was confined to beyond $\sim$5 AU, and its post-1974 $\sim$6 year orbit has a perihelion near the orbit of Mars. Internal heating from the decay of radionuclides is inefficient for such small and porous bodies as comets, where the radionuclide-containing refractory materials are heavily diluted with ice \citep{nkm11}. However, \citet{rei87} have suggested that aqueous alteration might occur near the surface of an active, contemporary comet when the temperature exceeds 200 K. 

Infrared spectra of material excavated from the surface of comet 9P/Tempel 1 by the Deep Impact mission could be fit with phyllosilicates and carbonates \citep{Lisse:2006gf,Lisse:2007mz}, but the fits are non-unique and do not provide strong proof of the presence of these phases in Tempel 1. No phyllosilicates or cabonates have been observed in ultra-carbonaceous Antarctic micrometeorites \citep{engrand2016variations}, or in comet 67P/C-G by the Rosetta mission \citep{rubin2019volatile}.

Craters on Wild~2's surface imaged during Stardust's encounter with the comet were speculated to be impact craters, but are now thought to be depressions caused by the release of volatiles from the nucleus \citep{thomas2008loss}. Comet nuclei imaged by spacecraft appear very different from asteroids: it is likely that small comets formed small and were not subjected to the collisional environments seen by asteroids (small asteroids appear to be fragments of larger bodies). This is consistent with the low density of impact craters on Pluto and Arrokoth \citep{singer2019impact,spencer2020geology}. The unusual bilobate structure of Arrokoth implies that it formed as a contact binary at very low speeds and was not disturbed by subsequent collisions \citep{mckinnon2020solar}. So it is unlikely that impact craters could have caused significant melting of water ice.

Secondary minerals have been identified in the returned Wild 2 samples, though they are scarce. \cite{ber11} reports Ni-, Cu-, and Zn-bearing iron sulfide assemblages in Track 26 (C2054,5,26,0,0) that may be the products of low-temperature aqueous alteration \citep{berger2015experimental}. The terminal particles of Track 26 are both approximately 10$\times$20~$\mu$m and made of crystalline SiO$_2$ (with up to 1 wt. \% Al$_2$O$_3$) nodules rimmed by 100~nm fayalite (Fa$_{96}$) with up to 6.6 wt.\ \% MnO \citep{joswiak2012comprehensive}. The Ada fragments are similar to rare silica-fayalite chondrules found in type 3 unequilibrated ordinary chondrites \citep{brigham1986silica}, though Ada is significantly more enriched in MnO. The extremely ferroan composition of the olivine in these particles is indicative of formation by interaction with water \citep{was94}, or, alternatively, by high-temperature processes \citep{brigham1986silica}. 

\citet{nguyen2017coordinated} report a 12~$\mu$m particle made of symplectically intergrown pentlandite and nanocrystalline maghemite in track 147 (C2101,1,147,0,0). The formation of mineralogically similar objects in the Acfer 094 meteorite is not fully understood, but likely involves interaction with liquid water \citep{matsumoto2022three}. TEM analyses by \citet{stodolna2012mineralogy} identified magnetites up to 1.5 $\mu$m in diameter from the bulb of track C2092,3,80 which made up $\sim$10\% of the total analyzed crystalline material. These small particles may have original made up a magnetite framboid like those seen in some carbonaceous chondrites \citep{hua1998unusual}. \citet{hicks2017magnetite} identified magnetite in two tracks: Track 178 (C2045,4,178,0,0), Track 187 (C2065,4,187,0,0), and three terminal fragments using Fe x-ray absorption spectroscopy. Magnetite is usually associated with more-abundant phyllosilicates in aqueously altered carbonaceous chondrites, and phyllosilicates are the dominant phase in chondritic-smooth IDPs \citep{bra96}. However, well-preserved phyllosilicates have not been identified in the Wild 2 samples. High-speed capture into aerogel could modify phyllosilicates, but even then the thermal breakdown products of phyllosilicates could be identified by TEM \citep{zol06}. Only one grain has been reported that is a possible dehydrated phyllosilicate \citep{sch11}.

The general paucity of secondary phases in Wild 2 indicate that parent-body aqueous alteration of the sort seen in CI, CM, and CR carbonaceous chondrites did not occur on the comet. Instead, the scarce secondary phases identified in the Stardust collection likely formed via aqueous alteration on an asteroid before they were disrupted via impact and then transported to the scattered disk to be eventually incorporated into Wild 2. Aqueous alteration in carbonaceous chondrite parent bodies occurred around 5 Myr after the onset of CAI formation \citep[e.g.,][]{fuj13} and possibly continued for a few million years as determined by ion probe measurements of the $^{53}$Mn--$^{53}$Cr system \citep{jil13}. Therefore, if these grains were transported to the scattered disk for incorporation into comet Wild 2, transport must have occurred relatively late. 

Thermal metamorphism in chondrites is assessed using a suite of samples from the same group with different metamorphic grade \citep{grossman2005onset}. With knowledge of the starting composition (type 3.0), it is possible to track diffusion of elements between phases as metamorphic grade increases (e.g., the diffusion of Cr out of olivine). With the comet Wild~2 samples, using a system like that of \citet{grossman2005onset} requires strong assumptions about the composition of starting materials using chondrite analogs. Additionally, the small grain size and fragmentary nature of Wild~2 samples makes diffusion lengths and timescales, at a given peak temperature, very difficult to determine compared to a meteorite sample. The low amount of Cr (compared to the most primitive 3.00 chondrites) in some Wild~2 olivines led \citet{frank2014olivine} to conclude that Wild~2 experienced mild metamorphism (metamorphic grade 3.05--3.15), though this estimate is very model-dependent. Cr contents of Wild~2 olivines are higher than the large majority of chondrite olivine, and the formation of Cr-rich rims on olivine via Cr diffusion is a very sensitive indicator of metamorphism, but none have been found in Wild~2. There is no evidence of strong thermal metamorphism in the Wild~2 samples. Zoning in larger individual terminal fragments may provide insight into thermal history, but hypervelocity capture can also induce chemical zoning \citep{Tomeoka:2008fy} and the size of Wild~2 grains is usually smaller than the zoning length scale \citep{gainsforth2015constraints}.

\section{Organics and Carbon}
\label{organics}
Analysis of organics in Wild~2 was a secondary goal of the Stardust mission and is complicated by contamination of the collecting medium and the effects of hypervelocity capture \citep{spencer2007comment,sandford2010assessment,de2011correlated,chan2020concerns}. The contamination problem necessitates the use of techniques with high spatial resolution (e.g., TEM, STXM, NanoSIMS, nanoFTIR) to distinguish native Wild 2 organics (associated with tracks) from contaminants. Initial reports from the Stardust preliminary examination of highly heterogeneous Wild 2 organics compared to chondrites and IDPs \citep{sandford2006organics} were likely due to organic contaminants. A reanalysis found that only a small fraction of these grains are unaltered cometary organic matter, and they are represented by one functional group profile similar to that identified in meteorites and IDPs \citep{de2011correlated}. \citet{cody2011establishing} also identified similar functional group concentrations between comet Wild 2, interplanetary dust, and chondritic IOM, and concluded the organics from these three sources all derived from a formaldehyde polymer. The amino acid glycine was detected in Stardust foils lining the aerogel \citep{elsila2009cometary}, its high $\delta^{13}$C value of $+29 \pm 6$\permil\ evidence of its extraterrestrial origin (glycine was also detected in comet 67P/C-G \citep{hadraoui2019distributed}). Carbonaceous materials in comet Wild 2 have morphologies and sizes very similar to those seen in IDPs, and somewhat similar to chondritic IOM \citep{matrajt2012diverse,matrajt2013origin,matrajt2013textures}. 

Organic nanoglobules with isotope anomalies in hydrogen and nitrogen are found in primitive chondrites \citep{de2013isotopic}. Two such globules have been found in the Wild~2 samples, and they are quite different. A 1.5~$\mu$m nanoglobule in track 80 (C2092,2,80,0,0) has normal hydrogen isotopes but $\delta^{15}$N = 1120\permil\ (higher than neighboring organic matter) and is highly aromatic. A 2.5~$\mu$m nanoglobule from track 2 (FC3,0,2,0,0) has $\delta$D = 1000\permil\ and normal nitrogen isotopes. The track 2 nanoglobule is made of non-aromatic organics and contains abundant nitrile and carboxyl functional groups (though the authors caution that the track 2 nanoglobule could be contamination: the C-XANES is very similar to cyanoacrylate, and electron exposure in the TEM may cause large D-enrichments in organics). A nanoglobule in Murchison has a similar N isotope anomaly and highly aromatic chemistry, similar to the track 80 nanoglobule \citep{de2013isotopic}. The authors conclude that Wild 2 sampled the same reservoir of organics that was sampled by the parent asteroids of chondritic meteorites.

The carbon content of comet Wild~2 is difficult to estimate precisely because of the destruction of labile carbon-bearing organics during capture, and percent-level carbon contamination in the aerogel. However, more refractory C-bearing phases like carbonates, carbides, and graphite would have survived aerogel capture. The C/Si ratio in comet Wild~2 is likely similar to the carbonaceous chondrites (Figure \ref{fig:C_to_Si}), corresponding to a few weight percent carbon, though it is not currently possible to say which meteorite group is closest to Wild~2 in carbon content. It is clear, however, that comet Wild~2 contains far less carbon than comet Halley \citep{zubko2012evaluating}: 25\% of Halley particles were nearly pure carbon (like the UCAMMs) and another 50\% were 10--90\% carbon \citep{fomenkova1992compositional}. Comet Wild~2 also likely has a much lower C/Si ratio than the Sun. If Wild~2 is as carbon-rich as Halley, entire tracks would be comprised of carbonaceous grains. Such carbon-rich tracks (and Al-foil impact residues) would be obvious even with carbon contamination in the collecting medium, and these are not observed.



\section{Kool assemblages}
\label{kool}
The large majority of minerals and mineral assemblages seen in the Stardust samples are also seen in meteorites sampled from asteroids. The most significant assemblage found in the Stardust samples that is not seen in asteroid samples is the Kool assemblage: kosmochloric high-calcium pyroxene and associated minerals \citep{joswiak2009kosmochloric}. The pyroxenes in Kool grains are usually augites with extremely high abundances of Cr$_2$O$_3$ (up to 12 wt. \%) and Na$_2$O (up to 6 wt.\ \%). Other minerals in Kool assemblages are Fe-rich olivine (Fo$_{70\text{--}85}$), albite or aluminosilicate glass, and sometimes Cr-rich spinel. About half of studied Stardust tracks contain Kool assemblages, and they are also found in chondritic-porous interplanetary dust particles \citep{joswiak2009kosmochloric}. Two Stardust Kool grains (T77-F1 and T77-F5) have oxygen isotopic compositions close to or slightly above the terrestrial fractionation line ($\Delta^{17}$O=1.3$\pm$1.7 and 1.8$\pm$2.5 (2$\sigma$), \citet{nakashima2012oxygen}).  A Kool grain from a giant cluster IDP, LT14, was measured for O isotopes by \citet{zhang2021oxygen}: $\Delta^{17}$O=1.1$\pm$1.6 (2$\sigma$). The oxygen isotopic composition of Kool grains is consistent with type II chondrules (Mg\#$<$90) in ordinary chondrites (measured to be $\Delta^{17}$O $\approx 0.5\permil$ by \citet{kita2010high}). In addition to the O isotopes, the mineralogy, elemental composition, and small-grain sizes of Kool grains support the hypothesis that they are the precursors of type II chondrules in ordinary chondrites.

Kool assemblages appear to be unique cometary rocks as they are found in interplanteary dust and the comet Wild 2 samples but not (in significant abundance) in meteorites. Similarly, filamentary enstatite and GEMS grains are found in IDPs but not in significant abundance in meteorites (Section \ref{ucammidpmeteors}). The effects of hypervelocity capture make it difficult to identify these two intriguing IDP phases in the Stardust samples \citep[e.g.,][]{ishii2008comparison}, though there have been reports of both \citep{stodolna2014characterization,gainsforth2019fine}.

\section{Reduced phases in comet Wild 2}
\label{reducedphases}
A study of 15 individual chondritic-porous interplanetary dust particles (CP-IDPs) and 194 Stardust fragments showed that Fe is more reduced in the Wild~2 samples compared to the CP-IDPs \citep{ogl10}. Metallic Fe is more abundant in the Wild~2 samples than individal porous IDPs, but when looking at all particles in a giant cluster IDP, metallic Fe and the total oxidation state of Fe is consistent with comet Wild 2 \citep{westphal2017oxidation}. An entire giant cluster IDP, which includes large silicates, sulfides, and other non-porous grains, is a more representative sample of cometary solids, and more similar to Wild 2 material, than individual CP-IDPs. The 10~$\mu$m $\times$ 15~$\mu$m kamacite grain Simeio from Track 41 (C2044,0,41,0,0) was analyzed for minor elements by LA-ICP-MS \citep{humayun2015preliminary}. This grain has lower Ni content ($\sim$2 wt.\ \%) than metal grains from chondrites. Simeio is most similar in Co, Ni, Ga, Ge and Ir content to ureilite metal. Ureilites are enigmatic, carbon-rich meteorites with uncertain origins \citep{goodrich2004ureilitic}. More trace-element analyses of kamacite and taenite in the Stardust and IDP collections is needed to better understand the origin of cometary metal.

\begin{figure}[!ht]
\begin{center}
\includegraphics[height=0.25\textheight]{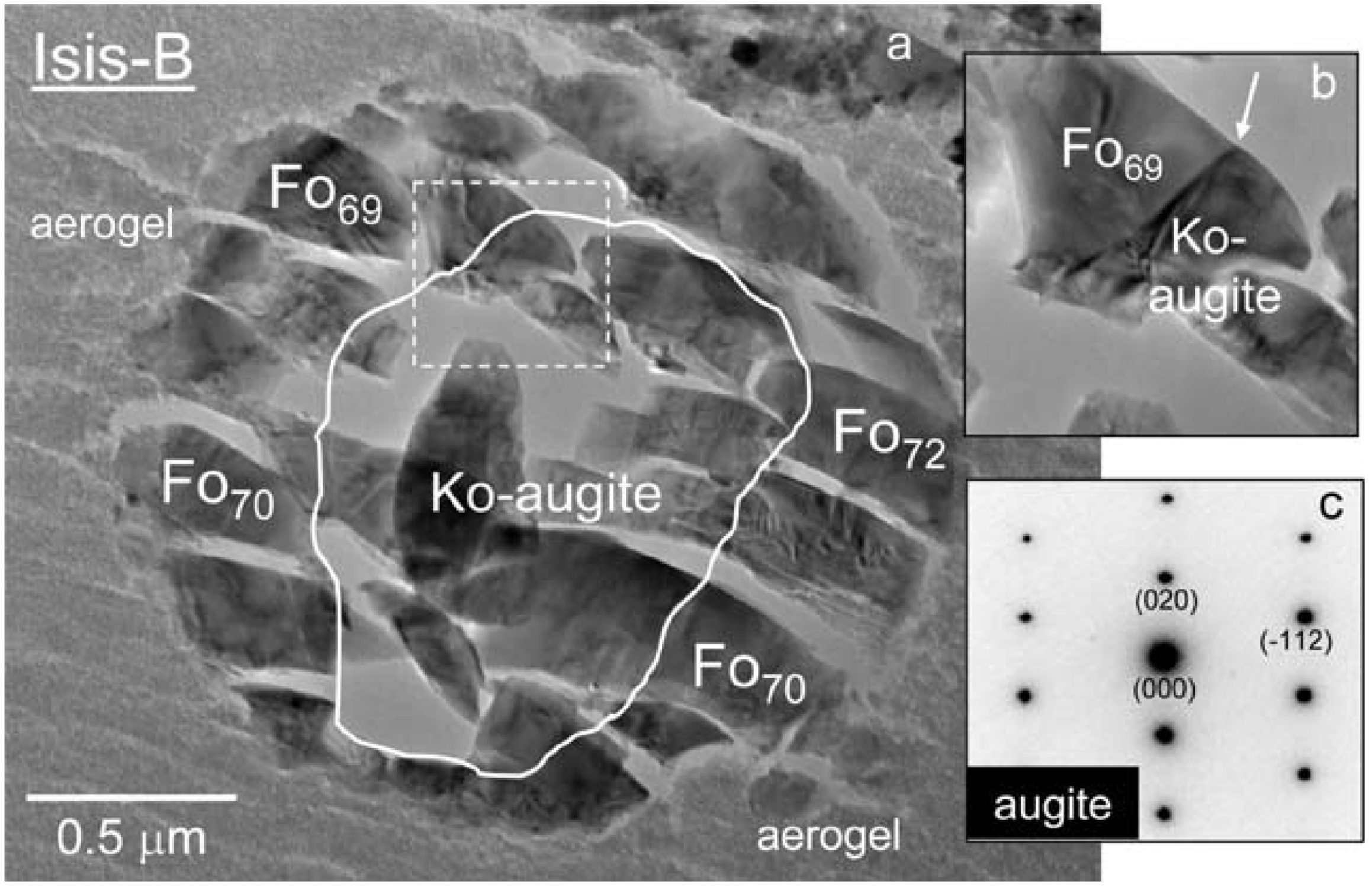}
\includegraphics[height=0.25\textheight]{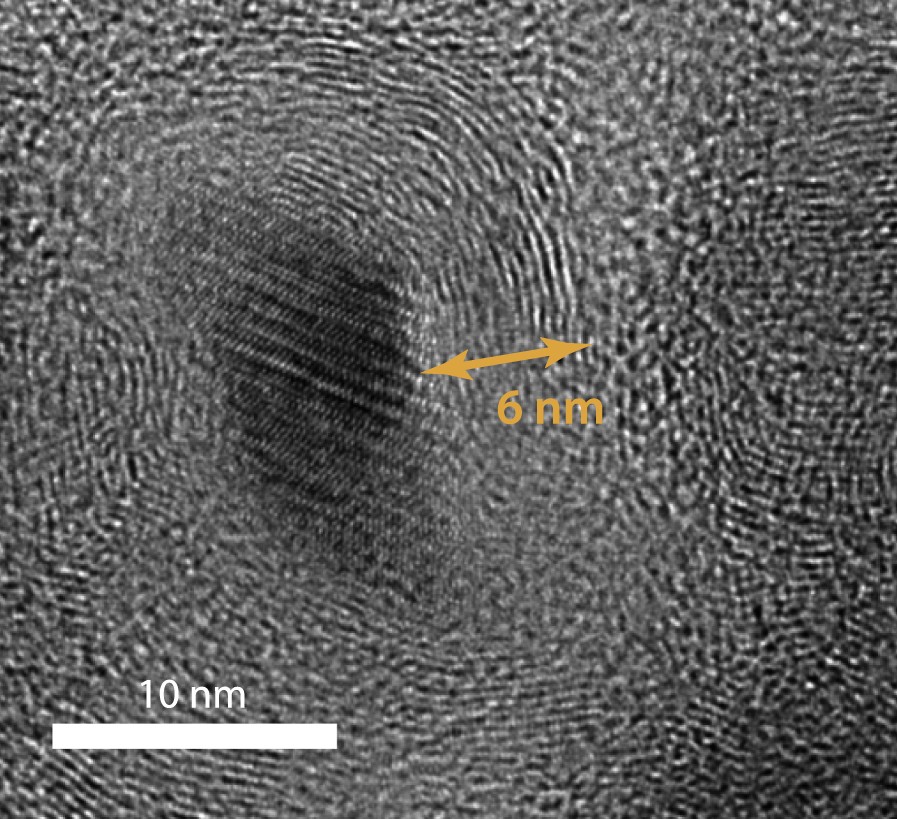}
\end{center}
\vspace*{-0.25in}
\caption{Left) a) Kool grain Isis-B from track 41, from \citet{joswiak2009kosmochloric}. White line is contact between Ko-augite and olivine. b) Sharp contact between olivine and Ko-augite. c) Selected-area electron-diffraction pattern obtained of augite on [201] zone axis. Right) HRTEM image of a 6-nm PGC shell around an Fe-carbide core. From \citet{de2017evidence}.  \label{fig:kool_and_pgc}}
\end{figure}

The comet Wild~2 samples include unusual material that does not have a direct analog in carbonaceous chondrites or interplanetary dust particles. \citet{de2017evidence} describe a 20~$\mu$m aggregate of nanoparticulate Cr-rich magnetite found in track 183 (C2103,24,183,0,0). The magnetite contains poorly graphitized carbon (PGC) that form shells (possibly by Fischer-Tropsch-type reactions) around 5--15~nm cores of highly reduced Fe carbide (Figure \ref{fig:kool_and_pgc}). The assemblage may have originated in a C-rich, highly reduced, and warm region of the inner nebula and was later oxidized in a more distant icy region in the outer Solar System \citep{de2017evidence}. Such objects were not incorporated into chondritic parent bodies in significant abundances.

\citet{frank2014olivine} report forsterite terminal fragments from track 61 (C2009,6,61,0,2) with extremely low amounts of FeO, MnO, Cr$_2$O$_3$, and CaO. Such highly reduced forsterite is compositionally consistent only with nearly FeO-free forsterite found in aubrites (enstatite achondrites). Aubrites likely formed in a very reducing environment in the inner Solar System \citep{keil2010enstatite}. T175/F1 (Section \ref{aoa}) is an $^{16}$O-rich forsterite studied by \citet{fukuda2021correlated} that also has very low FeO, MnO, Cr$_2$O$_3$, and CaO, implying a similar origin as the track 61 terminal fragments. The Kakangari meteorite contains $^{16}$O-rich matrix forsterites \citep{nagashima2015oxygen} but with higher MnO than T175/F1. Aubrites are $^{16}$O-poor \citep{clayton1984oxygen} and do not contain $^{16}$O-rich forsterites. The discovery of two $^{16}$O-rich Wild~2 enstatites \citep{defouilloy2017origin} provides another link to Kakangari, as $^{16}$O-rich enstatites are found in this meteorite and in an IDP \citep{utt2023diverse} but nowhere else to date. The origin of these highly reduced forsterites in Wild~2 is uncertain, though they likely came from a reduced and $^{16}$O-rich inner Solar System reservoir.


\section{Comet Wild 2 fine-grained material}
\label{fines}
Terminal fragments in Stardust aerogel tracks collected from comet Wild~2 appear to be mostly refractory, and likely formed in high-temperature events in the inner Solar System \citep{bro12,ogliore2012incorporation}. The terminal fragments, however, are larger and easier to analyze and are therefore over-represented in the studied Wild~2 sample. This may have resulted in a biased view of rocky material from comet Wild~2. Fine-grained material is spread over the bulb of tracks and intermixed with aerogel making it much more difficult to identify (e.g., distinguishing cometary material from compressed aerogel) and analyze. 

 \citet{stodolna2012mineralogy} studied a piece of the bulb of Stardust track 80 by TEM. They found that that the cometary material contained two populations of approximately equal abundance: 1) olivine and pyroxene grains ($>1$~$\mu$m in size); 2) smaller fine-grained material with Fe, Mg, and S concentrations close to CI chondrites. The authors conclude that Wild~2 is composed of two types of materials: large evolved grains (which are similar to the larger and well-studied terminal fragments), and fine-grained material with primitive chemistry. 

\begin{figure}[!htbp]
    \centering
    \includegraphics[height=0.28\textheight]{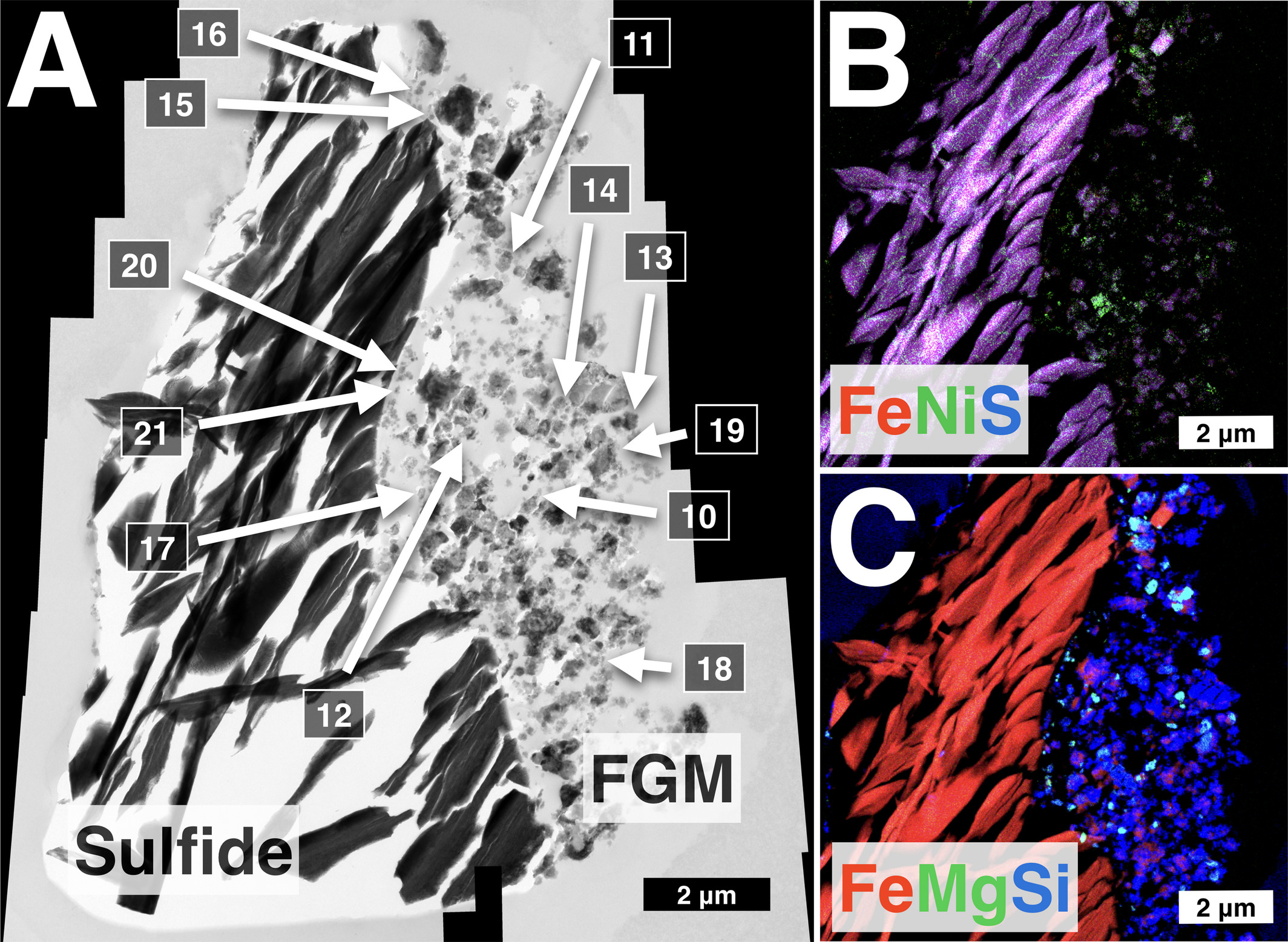}
    \includegraphics[height=0.28\textheight]{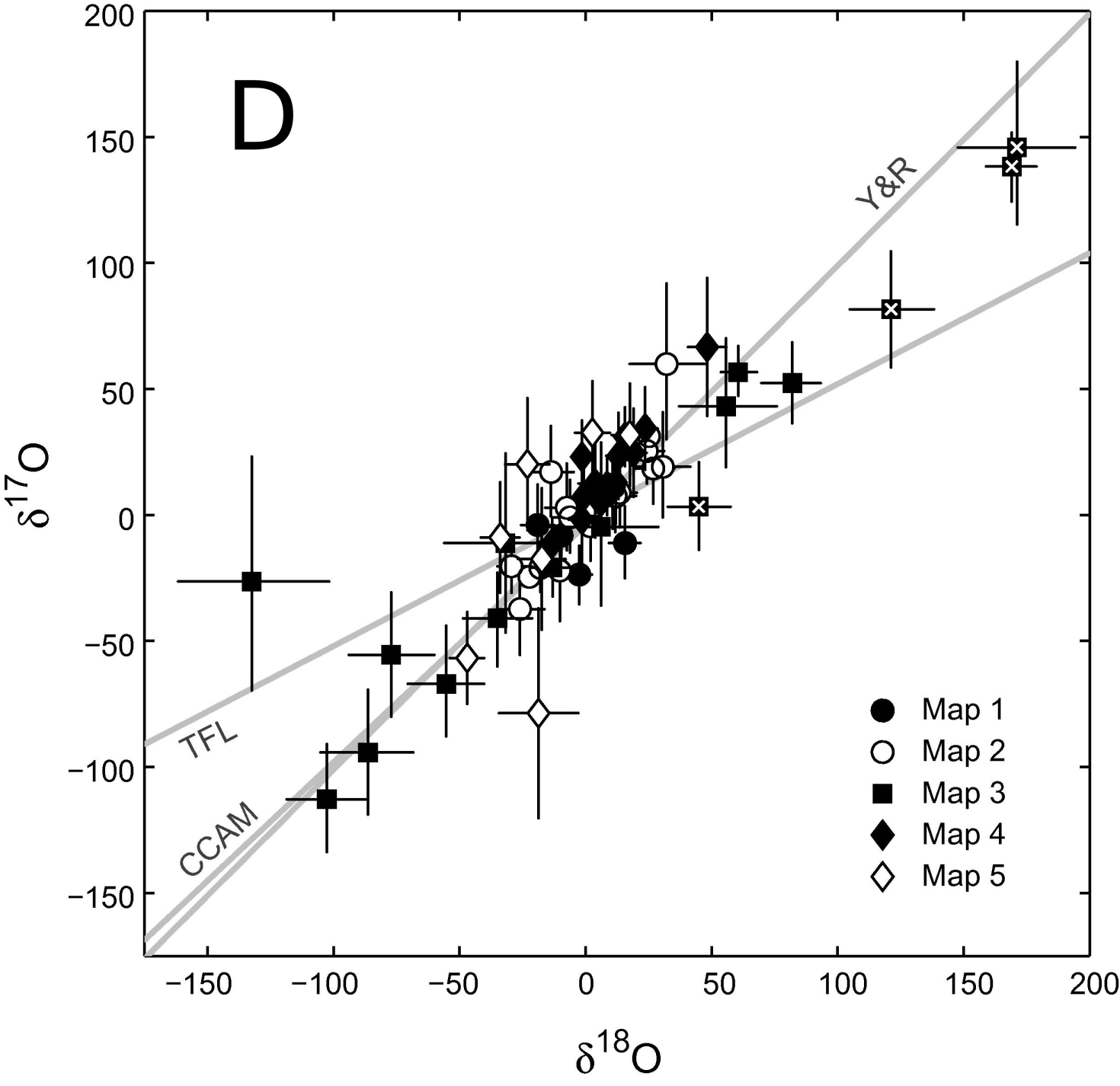}
    \caption{A) Bright-field TEM image of the Andromeda iron-sulfide grain (left) and fine-grained material (right). B) FeNiS RGB map. C) FeMgSi RGB map.  D) Oxygen 3-isotope plot showing the compositions of 63 particles in the bulb of Stardust track C2052,74. The terrestrial fraction line (TFL), carbonaceous chondrite anhydrous mineral line (CCAM), and Young \& Russell line (Y\&R) are shown. Uncertainties are 2$\sigma$. A--C are from \citet{gainsforth2019fine}; D is from \citet{ogliore2015oxygen}.
    \label{fig:fines}}
\end{figure}

Using ion probe techniques to analyze fine-grained material still embedded in aerogel, \citet{ogliore2015oxygen} found that the fines showed a very broad range of O isotopic compositions: $-70 \permil <  \Delta^{17}\text{O} < +60\permil$ (Figure \ref{fig:fines}). This study suggests that Wild 2 fines are either primitive outer-nebula dust or a very diverse sampling of inner Solar System compositional reservoirs that accreted along with a large number of inner-Solar-System rocks (the terminal fragments) and volatile ices to form comet Wild~2.

TEM studies of fine-grained material preserved near terminal particles have also shown that this material is distinct from the terminal fragments, and likely more primitive. \citet{gainsforth2019fine} studied fine-grained material in contact with a large terminal sulfide grain named Andromeda from track C2086,22,191 (Figure \ref{fig:fines}). Andromeda acted as a protective shield, preserving the fines mostly intact. This study, and another investigating fines associated with Iris \citep{stodolna2014characterization} (from track C2052,12,74) found that many of the fine particles were similar to grains found in CP-IDPs. \citet{jos12b} studied fine-grained material associated with Febo (track C2009,2,57) and identified $^{15}$N-rich organics which are thought to be a signature of primitive molecular cloud material \citep{keller2004nature}.

The fine-grained material in comet Wild~2 remains quite unknown due to alteration by hypervelocity capture, despite multiple efforts to investigate it over the last 18 years. The fines seem to have broadly chondritic composition and may be composed of: 1) GEMS grains, 2) unequilibrated aggregate mineral grains (like CP-IDPs), and/or 3) phyllosilicate-rich, hydrated matrix as is found in CI chondrites. All of these have approximately chondritic composition, but represent three distinct provenances of cometary fines. The identification of GEMS in the Wild~2 fines is challenging \citep{gainsforth2019fine,ishii2019comparison}, and phyllosilicates may have been destroyed by capture. Nonetheless, it is important to understand the mineralogy and origin of Wild~2 fines.


\section{Aluminum-26}
\label{26al}
The radioactive isotope $^{26}$Al has a half-life of 0.72 Myr and is an important heat source for early-formed bodies in the Solar System \citep{hevey2006model}. For example, the formation and evolution timescales of Enceladus and Iapetus depends sensitively on how much $^{26}$Al was incorporated into the forming satellites \citep{schubert2007enceladus,castillo200926al}. The initial abundance and homogeneity of $^{26}$Al in the Solar System is a critical parameter for planet/satellite formation and evolution. Assessments of the homogeneity of $^{26}$Al have been limited to asteroidal samples that likely do not sample the very outer parts of the Solar System. 

Several studies have argued for homogeneity of $^{26}$Al early in the Solar System's history. Many CAIs record the ``canonical'' amount of $^{26}$Al: initial $^{26}$Al/$^{27}$Al = 5.2$\times$10$^{-5}$ \citep{jacobsen2008}. Analyses of the Al-Mg system in 15 chondrules from the unequilibrated ordinary chondrite Semarkona suggest that $^{26}$Al was homogeneously distributed in the inner solar system where these chondrules formed \citep{villeneuve2009homogeneous}. High-precision Mg isotope measurements of olivine in primitive chondrites (formed outside the CAI reservoir) shows a 35 ppm deficit in $^{26}$Mg compared to the Earth, consistent with radiogenic ingrowth from a canonical initial $^{26}$Al/$^{27}$Al in the asteroid-forming reservoir \citep{gregory2020primordial}. Concordant $^{26}$Al-$^{26}$Mg and U-corrected Pb-Pb ages of achondrites when anchored to the early-formed andesite-like achondrite Erg Chech 002 supports a homogeneous distribution of $^{26}$Al among the meteorite reservoirs \citep{reger2023mg}.

Conversely, other studies have argued for $^{26}$Al heterogeneity. \citet{larsen2011} report heterogeneity in excess $^{26}$Mg in samples of chondrites and achondrites and argue that large-scale heterogeneity (up to 80\%) may have existed in the inner Solar System. Though the canonical ratio is common in CAIs, there are $^{26}$Al-poor and $^{26}$Al-rich CAIs with similar O isotope composition \citep{krot2012heterogeneous,makide2013heterogeneous}, implying a heterogeneous distribution of $^{26}$Al in the CAI formation region that was likely homogenized a short time later (before most chondrules formed).

To assess $^{26}$Al heterogeneity over the entire Solar System, it is necessary to consider samples from bodies that formed very far from the Sun. Most Jupiter-family comets are thought to originate from the scattered disk, a dynamical sub-class of the Kuiper belt \citep{duncan2004dynamical}. Observations and modeling of KBOs and comets can shed light on the possible presence of $^{26}$Al in the outer Solar System.  

The evidence against, and for, substantial $^{26}$Al in the outer Solar System is as follows. Some comets contain supervolatiles like CO ice, and significant thermal metamorphism from $^{26}$Al decay would drive off these compounds \citep{mckinnon2002initial}. The presence of amorphous water ice observed in comets implies an upper $^{26}$Al/$^{27}$Al limit of 4$\times$10$^{-9}$ after accretion, any more $^{26}$Al and heating will cause amorphous water ice to convert to crystalline ice \citep{prialnik1987radiogenic}. Accretion timescales of KBOs modeled by incremental growth are much longer than the half-life of $^{26}$Al (10--30 Myr to build KBOs with 10-100 km radii) \citep{kenyon2008formation}. However, models that allow for gas-drag to speed up accretion can build the same size objects in less than 1 Myr \citep{weidenschilling1997origin,weidenschilling2004icy}, and the timescale of inner Solar System grain migration to the outer Solar System is much shorter than the $^{26}$Al half-life \citep{boss2008mixing}. Even in small KBOs, $^{26}$Al heating will differentiate the body into a rocky core and icy mantle \citep{mckinnon2002initial}. Icy surfaces have been detected on KBOs \citep{brown2012water}, though no pure-ice comets (collisional fragments of a rock-from-ice differentiated KBO) have been observed. The diversity of KBO compositions and densities may require heating from $^{26}$Al \citep{mckinnon2008structure}.

Sample analysis is critical to help resolve the debate on the availability of $^{26}$Al in the outer Solar System. To date, the Al-Mg system has been measured in four Stardust particles: Coki, Inti, Iris, and Pyxie. None of these particles have a resolved initial $^{26}$Al/$^{27}$Al ratio and so only an upper bound is meaningful. In the following, a one-sided 2$\sigma$ (95.45\%) upper bound is stated (equal to the mean initial $^{26}$Al/$^{27}$Al plus 1.69 multiplied by the reported 1$\sigma$ uncertainty), which is smaller than the two-sided 2$\sigma$ upper bound (as may be reported in the original publication). 

The measurements made by \citet{matzel2010coki} of Coki (CAI fragment, described in Section \ref{chondruleaoacai}) correspond to a one-sided 2$\sigma$ upper limit of $^{26}$Al/$^{27}$Al = $8.2 \times 10^{-6}$ when Coki crystallized. The same bound for Inti (CAI fragment described in Section \ref{chondruleaoacai}) is $5.0 \times 10^{-5}$ \citep{ishii2010Inti} (very close the canonical value, so not a useful constraint). Iris is likely a type-II chondrule fragment \citep{gainsforth2015constraints} and was measured to have an upper bound of $3.0 \times 10^{-6}$ \citep{ogliore2012iris}. Pyxie is  made of FeO-poor, low-Ca pyroxene and plagioclase \citep{nakashima2015late}. The correlation between Pyxie's O isotopic composition and Fe/Mg ratio led the authors to conclude it is most similar to chondrules in CR3 chondrites. Pyxie's upper bound is $3.2\times 10^{-6}$ \citep{nakashima2015late}, which is similar to the upper bound on Iris.

The evidence for only dead $^{26}$Al in cometary samples is supported by work on interplanetary dust. The first measurements of Mg isotopes in interplanetary dust particles by \citet{esat1979magnesium} (using thermal ionization mass spectrometery) found excesses of $^{26}$Mg in three of four measured chondritic-porous particles. Though the magnitude of the anomaly was small, 3--4$\permil$, it was significantly larger than the uncertainty ($\sim$1$\permil$, 2$\sigma$). If this anomaly was due to the decay of $^{26}$Al, the initial $^{26}$Al/$^{27}$Al in these particles would be unrealistically high: $10^{-2}$. Using a Cameca IMS-3f ion probe, \citet{mckeegan1985ion} measured three IDPs by ion probe and found no anomalies in Mg isotopes. Subsequent ion probe measurements of two refractory particles also found no significant Mg isotope anomalies, with upper limits of 10$^{-6}$ and 7$\times$10$^{-6}$ for initial $^{26}$Al/$^{27}$Al \citep{mckeegan1987oxygen}. Two IDPs have been measured using modern ion probes, and both have no resolved initial $^{26}$Al. \cite{ogliore2020manchanito} reports an upper bound of $2.2 \times 10^{-6}$ for an unusual amorphous refractory particle from a giant cluster IDP of probable cometary origin. \citet{iskakova2023measurement} measured a CAI-like fragment from a giant cluster IDP and found an upper bound of $2\times 10^{-5}$.

 The youngest crystallization age of a comet Wild~2 component sets the earliest possible accretion age of comet Wild~2 (as the comet must have accreted after its last constituent grain crystallized, \citet{schrader2022prolonged}). The current lowest upper bound of initial $^{26}$Al/$^{27}$Al is in Iris, $3.0 \times 10^{-6}$, which corresponds to an age of CAI+3~Myr. Therefore, with the assumption that $^{26}$Al was initially homogeneously distributed in the young Solar System at the canonical level, Wild~2 must have accreted more than 3~Myr after CAIs. 

Alternatively, the $^{26}$Al-poor cometary grains may have formed before injection of $^{26}$Al into the solar nebula \citep{ouellette2007interaction}, though the probability of a supernovae occuring within 2~pc of the solar nebula in the first Myr of its lifetime is $<$1\% \citep{ouellette2010injection}. The low values of initial $^{26}$Al seen in some CAIs \citep{krot2012heterogeneous} can be explained if these objects formed before $^{26}$Al was injected. An early formation of the most refractory particles like hibonite aggregates \citep{koop2016link} seems plausible, but the large majority of ferromagnesian chondrules in chondrites formed at least 1~Myr later \citep{pape2019time}. The low $^{26}$Al in chondrule-like Wild~2 particles like Iris \citep{ogliore2012iris}, and other cometary grains with less refractory mineralogy than CAIs, likely reflects formation after $^{26}$Al had decayed.


\section{Discussion}
\subsection{Summary of main points}
 In Sections \ref{phasesidentified}--\ref{26al}, I have described representative laboratory analyses reported in more than 100 papers in the literature since the return of the Wild 2 samples, and included those that are, in my view, the most significant. Some important conclusions derive from the lack of definitive reports in the literature of important phases (e.g., phyllosilicates). The potential for selective alteration or destruction of certain critical phases adds to the challenge of interpretation (Section \ref{statanec}). Here I summarize the main conclusions from each section:

\begin{description}
    \item[Section \ref{phasesidentified}] Ferromagnesian silicates are the dominant phase in the rocky component of comet Wild~2, and carbonaceous phases are rare. 
    \item[Section \ref{caifragments}]  Refractory objects are found in Wild~2 in similar abundances as in carbonaceous chondrites, though the most refractory phases are not found in Wild 2.
    \item[Section \ref{minorelementsinsilicates}] Minor elements in ferromagnesian silicates show no clear correlations, which is similar to silicates in CR chondrites but unlike other chondrite groups.
    \item[Section \ref{oisotopes_o16rich}] Wild~2 contains a higher abundance of $^{16}$O-rich grains than other chondrite groups, except for CI chondrites and asteroid Ryugu, which have similar abundance.
    \item[Section \ref{oisotopes_o16rich}] Oxygen isotopes in ferromagnesian silicates from Wild~2 show a trend with Mg\# that is similar to CR chondrite chondrules.
    \item[Section \ref{psg}] Presolar grains appear to be relatively rare in comet Wild 2, though with limited statistics, very high abundances compared to chondrites cannot currently be ruled out.
    \item[Section \ref{aqueous}] Particles that are products of aqueous alteration or thermal metamorphism are rare in comet Wild~2.
    \item[Section \ref{organics}] Organics in comet Wild~2 appear to be of similar type and abundance as those found in carbonaceous chondrites. Total carbon content in Wild~2 is also likely similar to carbonaceous chondrites, and much lower than comet Halley and the Sun.
    \item[Section \ref{kool}] Comet Wild 2 (and cometary IDPs) contain at least one mineral assemblage (Kool assemblages) that were not accreted by asteroids sampled in the meteorite record.
    \item[Section \ref{reducedphases}] Unusual particles that contain highly reduced phases are found in the Wild~2 samples.
    \item[Section \ref{fines}] Fine-grained comet Wild~2 material is heterogeneous, broadly chondritic, and may represent unprocessed, primordial Solar System dust.
    \item[Section \ref{26al}] Though statistics are limited, it appears that comet Wild~2 has low levels of $^{26}$Al, probably because the grains accreted by Wild 2 formed after $^{26}$Al had decayed. Wild~2 likely accreted more than 3~Myr after CAIs.     
\end{description}

\subsection{Primordial inheritance of comet Wild~2}
The expected composition of primordial material is described in Section \ref{primordial}. In the comet Wild~2 samples, refractory carbonaceous material is very rare and the overall C/Si in comet Wild~2 solids seems to be quite low. Surviving organics appear similar to those found in carbonaceous chondrites. Amorphous silicates are rare compared to crystalline silicates and circumstellar grains are about as abundant as they are in asteroidal material (though uncertainties are large). This implies that the inherited interstellar material is similar in both abundance and nature as that found in carbonaceous chondrites from asteroids.

The primordial, interstellar component may be better represented in the finest-grained material that did not survive capture as well as the terminal fragments. The fines are far more difficult to analyze, and are underrepresented in the studied Wild~2 material. The smallest Wild~2 grains may contain a higher abundance of presolar (interstellar and circumstellar) grains. Tiny presolar grains mixed with melted-quenched aerogel and scattered in the three-dimensional bulbs of aerogel tracks are very difficult to identify. Labile organics, bearing light isotope anomalies from low-temperature reactions in the Solar System's parent molecular cloud, may have also been preferentially lost in hypervelocity capture. 

The comet Wild~2 samples can be compared to dust in forming planetary systems elsewhere in the Galaxy. About 4000 young-stellar objects (YSOs) are within 500~pc of the Solar System and can be observed astronomically in the infrared (5--40~$\mu$m) with the Spitzer Space Telescope \citep{watson2009mineralization}. Though there are large deviations, Spitzer observations of protoplanetary disks show some common trends in dust composition. Fe-bearing pyroxenes, olivines, sulfides, and oxides have not been definitively observed in the protoplanetary disks associated with YSOs, but are very abundant in comet Wild~2. However, Fo$_{90}$ olivine may be a better fit to the spectra of dust in protoplanetary disks than pure forsterite. Pyroxene is common in the Spitzer data and in Wild~2. Silica is also common in disk observations, but was only identified in the Ada fragments (Section \ref{aqueous}) from Wild~2. Protoplanetary disks and Wild~2 are deficient in phyllosilicates and carbonates. Iron-bearing amorphous silicates are common in protoplanetary disks, and a minor fraction ($\sim$5\%) of disks contain dust that is almost entirely small amorphous silicates. Small amorphous silicates (e.g., GEMS) may be present in the Wild~2 samples, but hypervelocity capture makes their identification challenging. The mineralogy of Wild~2 can be precisely determined, but there are significant challenges with linking infrared observations of distant astronomical objects to actual rock samples. Overall, the Wild~2 samples are generally consistent with disk observations (lack of secondary minerals, abundant Mg-rich crystalline silicates), but there are important differences (lack of Fe-bearing minerals in disk observations), and uncertainties (amorphous silicates). Future observations of protoplanetary disks with the James Webb Space Telescope will help us understand the differences and similarities between the solar nebula (as sampled by primitive bodies like comets) and other protoplanetary disks.

\subsection{Solar System inheritance of comet Wild~2}

Comet Wild~2 may have a similar amount and type of inherited non-icy interstellar material as meteorites derived from carbonaceous chondrites, but otherwise they are very different. The OSIRIS-REx and Hayabusa2 samples returned from carbonaceous asteroids (Bennu and Ryugu) are dominated by secondary phases that formed inside asteroids \citep[e.g.,][]{dionnet2023three}. Unlike other primitive meteoritic materials, the cometary samples are almost totally anhydrous and never resided in an ice-containing parent body with high enough internal pressure and temperature (driven by $^{26}$Al-heating) to retain liquid water. The Wild~2 solids and CP-IDPs are perhaps the only representative samples of the nebular solids that filled the outer Solar System. The larger Wild~2 fragments are mostly igneous rocks that formed in the Solar System, similar to objects found in asteroids. Is this igneous component from one reservoir, or many?

The $\Delta^{17}$O vs.\ Mg\# trend (Figure \ref{defouilloy_O_isotopes}), Mn and Fe contents in olivines (Figure \ref{Schrader_olivine}), and late formation of Pyxie and Iris ($>$3 Myr after CAIs)  implies that chondrule-like objects in Wild~2 may have originated from the same reservoirs as CR chondrite chondrules (which have an estimated age of 3.8 Myr after CAIs, \citet{tenner2019extended}). However, type I chondrules outnumber type II chondrules in CR chondrites by a factor of $\sim$25 \citep{schrader2013formation} and show a strong abundance peak in low-FeO olivine \citep{frank2014olivine}---both observations contrast sharply with the Wild~2 samples. Likewise, other similarities with objects in other chondrites do not apply to all compositional, chemical, or structural characteristics, or statistically over a large number of Wild~2 grains. The Wild~2 materials are not consistent with a single chondrite group, and are also not consistent with a weighted average of known meteorite types. Wild~2 (and CP-IDPs) appear to be unequilibrated aggregates of components that formed in multiple nebular environments.

Comet Wild~2 contains unique or very unusual particles compared to what is found in meteorites (Kool assemblages, metal grains of unusual composition, a carbon-magnetite nanoparticle assemblage, $^{16}$O-rich enstatite, a silica-fayalite assemblage). The presence of these phases in the Stardust samples implies that Wild~2 accreted dust that was not incorporated into meteorite parent bodies in significant amounts. Wild 2 likely sampled early generations of Solar System silicates that are not common in meteorites: abundant $^{16}$O-rich relict grains and Kool assemblages (type II OC chondrule precurors).

Wild 2 however is also remarkable for what it does not contain: significant amounts of hydrated phases or secondary phases that formed by the interaction of rock and water, material that shows effects of significant thermal metamorphism, or abundant achondrite material. The large majority of chondrites in the meteorite record show evidence of alteration by heat or water. Only a handful of the $>$64,000 known chondrites are petrologic type 3.0. If Wild 2 formed from the disruption of already-formed chondrite parent bodies, it should have incorporated the altered material that is sampled by meteorites found on Earth. The fact that Wild~2 contains mostly material unaltered on a previous parent body implies that it directly accreted dust that formed in the solar nebula, but only minor amounts of impact debris from disrupted asteroids. 

Dust was accreted by Wild~2 as late as CAI+3~Myr (assuming the Al-Mg system in Iris can be used for age dating), which is after the accretion of the ordinary chondrite and most carbonaceous chondrite parent bodies \citep{sugiura2014correlated}. As the solar nebula began to disperse, the outward transport mechanisms that served as a conveyor belt to the outer Solar System subsided. The time that the solar nebula transitioned to a debris disk (devoid of dust and gas) can be estimated by the crystallization ages of chondrules in (matrix-free) CB chondrites (thought to form in an impact plume generated by colliding planetesimals). \citet{bollard2015pb} measured Pb-Pb ages in chondrules from the CB$_\text{a}$ chondrite Gujba to be 4.8$\pm$0.3 Myr after CAIs. This is likely close to the time when the nucleus of Wild~2 completed its accretion. The debris from asteroids (including achondrite parent bodies) disrupted around this time or later would not be accreted by Wild~2.

\subsection{Wild~2 on the asteroid-comet continuum?}

It has been proposed that comets and asteroids exist on a continuum based on, for example, objects that appear to have asteroid-like orbits but cometary activity \citep{hsieh2017asteroid}. A ``continuum object'', an asteroid with cometary activity, may show mineralogy consistent with meteorites we have in our collections (derived from asteroids). Comet Wild~2 is not such an object as its mineralogy is distinctly not asteroidal \citep{bro12}. Asteroids accreted mostly ``local'' materials that were subsequently processed by heat and water (to varying degrees) on the asteroid which lead to further equilibration and homogenization of the accreted components. Asteroidal material has distinctive properties (e.g., chondrule sizes) which allow us to group together different meteorites that fall to Earth at different times and places. Comet Wild~2 contains dust from different regions and conditions in the Solar System which was not significantly modified after it accreted into Wild~2's nucleus. In this way, comet Wild~2 does not exist on a continuum with asteroids---it is a completely distinct Solar System body in its origin, evolution, and composition.





\subsection{Solar System dichotomy and comet Wild 2}
Isotope analyses of meteorites show that the sampled Solar System bodies tend to fall into two clusters when different isotope ratios (measured on the bulk sample) are plotted against each other: the NC (non-carbonaceous) and CC (carbonaceous chondrite) groups \citep{scott2018isotopic}. The most popular current explanation of the dichotomy is that a forming Jupiter opened a gap in the protoplanetary disk, which kept the inner and outer Solar System reservoirs separate, preserving the isotopic imprint of heterogeneous infall from the molecular cloud \citep{kleine2020non}. Gaps possibly created by forming planets are seen astronomically in protoplanetary disks around other stars \citep{long2018gaps}. Alternatively, the dichotomy can be explained by a migrating snowline that results in two separate populations of planetesimals, that then grew through collisions to form larger bodies \citep{lichtenberg2021bifurcation}. The dichotomy may also be explained by selective destructive of presolar carriers of the isotope anomalies \citep{trinquier2009origin}. The Jupiter-gap scenario posits a siloed young Solar System when the chondrite parent bodies are accreting: communication between the inner and outer Solar System is cut off to preserve the isotope dichotomy we see in the meteorite record. 

The comet Wild~2 samples tell are remarkably different story. Migration of dust from many different regions of the inner Solar System (e.g., reduced phases, ordinary-chondrite-like Mn and Fe signatures in olivine) and outer Solar System (e.g., CR-chondrite like oxygen isotope signatures) combine to form the rocky component of Wild~2's nucleus. 

The efficient migration of inner Solar System material tens of AU radially outward through the disk midplane \citep{bockelee2002turbulent,ciesla2007outward} could have occurred before Jupiter opened a gap in the disk. However, measurements of meteorites have shown that the isotope dichotomy existed very early ($<$1~Myr), requiring Jupiter's rocky core to form $\sim$0.6~Myr after CAIs \citep{desch2018effect}. At least some of the particles in Wild~2, such as Iris and Pyxie, were likely transported outward long after this as their Al-Mg ages are at least $\sim$3~Myr after CAIs.

Iris, Pyxie, and other small Wild~2 grains can be transported outward through the Jupiter gap because it is a very porous barrier to small particles \citep{weber2018characterizing}. Grains smaller than a few hundred microns can efficiently pass through the Jupiter gap \citep{schrader2022prolonged}. The comet Wild~2 grains returned by Stardust are much smaller than this, meaning that the Jupiter gap could censor the coarse-grained components of the NC and CC asteroidal reservoirs, but allow comets to accrete fine-grained dust from all over the solar nebula.



\subsection{Comet Wild 2 recorded the Solar System's wild youth}
Crystalline material is observed astronomically in the outer reaches of protoplanetary disks \citep{juhasz2010dust,de2012comet,sturm201369}, well beyond the ``crysallization front'' of forsterite, the radial distance from the star where the disk temperature falls below the temperature at which amorphous silicate dust will crystallize (600--1000~K for forsterite, \citet{hallenbeck2000evolving}). Various disk transportation mechanisms have been proposed to explain the movement of crystalline material to the outer reaches of protoplanetary disks \citep[e.g.,][]{arakawa2021crystallinity}.  The crystalline dust in Stardust is not annealed dust, it is high-temperature igneous dust, so theories to explain crystalline dust by annealing (e.g., \citet{ali2023dust}) do not explain the comet Wild~2 samples.

The lack of a significant contribution of molecular cloud material (carbonaceous material, organics with light-element isotope anomalies, circumstellar and interstellar dust) in the comet Wild 2 samples implies that the non-volatile solids at the time of Wild~2's accretion were dominated by this transported Solar System material. That is, comet Wild 2's level of ``interstellar inheritance'' \citep{alexander2017measuring} is lower than expected for a comet. However, this may not be true for other comets that formed in a different time and place, or even for another part of comet Wild~2's nucleus. The processes driving outward transport of material in the solar nebula may have varied greatly on million-year timescales \citep{armitage2001episodic}, and comet Wild 2 may have accreted over millions of years \citep{mckinnon2008structure}. This means that different parts of Wild~2's nucleus could have different relative proportions of Solar System and molecular cloud component. Comet C/2017 US$_{10}$ (Catalina) showed an increase in the amorphous carbon to silicate ratio between two epochs separated by only six weeks \citep{woodward2021coma}. 

There is observational evidence that Wild~2 is not the only comet that has a comparable level of interstellar inheritance as chondrites. Two sun-grazing comets, Comet Lovejoy (C/2011 W3) and C-2003K7 (both from the Kreutz family of comets, and possibly derived from the same progenitor), show C/Si ratios consistent with chondrites \citep{ciaravella2010ultraviolet,mccauley2013extreme} (Figure \ref{fig:C_to_Si}).

The surprising result of the Stardust mission is that Wild~2 is not made of stardust, but is made of a diverse collection of Solar System dust. Instead of a time-capsule record of the primordial building blocks of the Solar System, Wild~2 is a time-capsule record of the turbulent formative years of the Solar System. During this time (from CAI formation to $\sim$5~Myr later), many processes may have shaped the forming Solar System: the Sun accreted material and may have underwent outbursts that heated portions of the solar nebula, chondrules formed via a number of exotic energetic processes, turbulence in the disk transported high-temperature inner Solar System rocks to the Kuiper belt, Jupiter's rocky core separated the inner from the outer Solar System, planetesimals formed, and giant planet migration wreaked havoc on the inner Solar System. Further analyses of the Wild~2 samples may yield insights into these important early Solar System events.











\subsection{Future directions} 

\subsubsection{More analyses of Wild~2 material}

Even after 18 years and more than 100 papers, we still have analyzed only a small fraction of the collected Wild~2 material. There are many large tracks in the aerogel collector that have yet to be harvested and analyzed. To understand a complicated body like comet Wild~2, it is important to analyze as much material as possible. Continued analyses of Wild~2 material, particularly TEM studies and O isotope analyses, will help us answer some of the questions presented here. Preparing and analyzing Wild~2 samples is challenging and requires specialized skills. Training the next generation of scientists in small-particle handling and analysis \citep{draper2017coordinated} is critical for the future of cometary dust science.

\subsubsection{Studies of primordial material}
The comet Wild~2 samples analyzed so far are not a collection of primordial building blocks as expected. However, primordial material (Section \ref{primordial}) can be found in a variety of extraterrestrial materials, including samples where the inclusion of such material is surprising. Carbonaceous clasts in the Isheyevo meteorite (a hybrid CH-CB chondrite, 50--70 vol \% Fe-Ni metal) show the largest $^{15}$N anomalies measured in samples or observed elsewhere in the Galaxy \citep{furi2015nitrogen}. The ungrouped carbonaceous chondrite Acfer 094 contains cosmic symplectite, a phase that recorded O and S isotope anomalies indicative of photochemical processing of the protosolar nebula by nearby stars \citep{vacher2021cosmic}. \citet{nittler2019cometary} identified a carbon-rich clast in the CR2 chondrite LaPaz Icefield 02342 which contains presolar grains, extremely $^{16}$O-rich material, and possibly GEMS grains. CI chondrites and asteroid Ryugu do not have primitive mineralogy (they are aqueously altered) but have primitive (solar) chemical compositions \citep{lodders2021relative,ito2022pristine} and contain clasts rich in carbon and presolar grains \citep{nguyen2023abundant}. UCAMMs carry N-rich organics, possibly of interstellar origin, unique among extraterrestrial materials \citep{dartois2018dome}. Finally, the fine-grained Wild~2 material was severely modified by aerogel capture, but it may be more primordial than the larger fragments. For example, the Wild~2 fines may be rich in GEMS and presolar grains. Our current collection of extraterrestrial materials likely includes undiscovered records of the Solar System's primordial building blocks.

\subsubsection{Future Cometary Samples}
Comets are a very diverse population of Solar System bodies---their volatile contents, D/H ratio, C/Si ratio, and other characteristics vary enormously from one comet to another. This may be because cometary nuclei accreted at different times and places in the early Solar System. The rocky components from comets may be highly variable as well. However, CP-IDPs (which are probably samples of short-period comets) are quite similar to Wild~2 (for grains larger than 2~$\mu$m). CP-IDPs are complex---they are mixtures of silicates, sulfides, amorphous phases, and organics, but they are relatively uniform. For example, no CP-IDPs are made entirely of isotopically anomalous presolar grains, amorphous silicates, or crystalline silicates that formed by annealing of glassy precursors. Because they are mixes of materials formed in numerous distant environments, it is possible, and perhaps likely, that large portions of the outer Solar System may have received similar mixes of inner Solar System materials. It is possible that most comets contain similar mixes of solid grains even though their volatiles differ. If chondrule fragments made it to Wild's formation region they must have gone everywhere, at least over some limited time period. Collecting samples from as many comets as possible through targeted stratospheric dust collections  \citep{messenger2002opportunities} is a relatively inexpensive way to sample from a diverse comet population and test this hypothesis.

The Planetary Science and Astrobiology Decadal Survey 2023--2032 recommends a (non-cryogenic) comet surface sample return mission for NASA's New Frontiers program. Such a mission would capture a gram-sized sample from a comet nucleus at low speed and return it to Earth for lab analyses. A low-speed sample capture would avoid the hypervelocity-capture degradation experienced by the Stardust samples, which compromised the organics and other phases, and the finest-grained rocky material. A gram-sized (or larger) sample mass would allow for bulk, high-precision isotope analyses of cometary solids, and identification of rare interstellar material. 

A cryogenic comet surface sample return would return ice as well as rocky material from a comet's nucleus \citep{bockelee2021ambition,westphal2021cryogenic}. Since cometary ices are almost certainly primordial remnants of the Solar System's molecular cloud, laboratory analyses of cometary ices would have tremendous scientific value. 

\section{Acknowledgements}
I gratefully acknowledge the reviews by three anonymous reviewers on an earlier version of the manuscript. Additionally, I thank Astrid Holzheid, Andrew Westphal, Gary Huss, Don Brownlee, Jeff Hester, and Klaus Keil for helpful discussions related to this manuscript. This work was supported in part by NASA grants NNX14AF22G and NNX12AF40G to RCO.

\bibliographystyle{model4-names}
\newpage
\bibliography{references}
\end{document}